\documentclass[twocolumn]{aa}
\usepackage{graphicx}
\usepackage{amssymb}
\usepackage{epstopdf}
\usepackage{upgreek}
\usepackage{lscape}
\usepackage[locale=US]{siunitx}

\begin{document}

\title{Dust depletion of Ca and Ti in QSO absorption line systems}
 
\author{
 C. R. Guber \inst{1},
 \and
 P. Richter \inst{1,2}
}

\offprints{C. R. Guber\\
\email{guber@astro.physik.uni-potsdam.de}}

\institute{Institut f\"ur Physik und Astronomie, Universit\"at Potsdam,
           Karl-Liebknecht-Str.\,24/25, 14476 Golm, Germany
\and
Leibniz-Institut f\"ur Astrophysik Potsdam (AIP), An der Sternwarte 16,
14482 Potsdam, Germany
}

\date{Received xxx; accepted xxx}

%%%%%%%%%%%%%%%%%%%%%%%%%%%%%%%%%%%%%%%%%%%%%%%%%%%%%%%%%%%%%%%%%%%%%%%%%%%%%%%%%%%%%%%%%%%%%%%%%%%

% \abstract{}{}{}{}{} 
% 5 {} token are mandatory
 
\abstract
{}
{To explore the role of titanium- and calcium-dust depletion in
gas in and around galaxies we systematically study Ti/Ca abundance ratios in intervening 
absorption-line systems at low and high redshift.}
{We investigate high-resolution optical spectra obtained by the UVES instrument at the 
{\it Very Large Telescope} (VLT) and spectroscopically analyze 34 absorption-line systems 
at $z\leq 0.5$ to measure column densities (or limits) for Ca\,{\sc ii} and Ti\,{\sc ii}.
We complement our UVES data set with previously published absorption-line data on Ti/Ca 
for redshifts up to $z\approx 3.8$. Our absorber sample contains 110 absorbers 
including Damped Lyman $\alpha$ systems (DLAs), sub-DLAs, and Lyman-Limit systems (LLS).
We compare our Ti/Ca findings with results from the Milky Way and the Magellanic Clouds 
and discuss the properties of Ti/Ca absorbers in the general context of quasar 
absorption-line systems.}
{Our analysis indicates that there are two distinct populations of absorbers with 
either high or low Ti/Ca ratios with a separation at $\left[ \mathrm{Ti}/\mathrm{Ca}\right]\approx 1$.
While the calcium dust depletion in most of the absorbers appears to be severe, the titanium 
depletions are mild in systems with high Ti/Ca ratios. The derived trend indicates that absorbers 
with high Ti/Ca ratios have dust-to-gas ratios that are substantially lower than in the Milky Way.
We characterize the overall nature of the absorbers by correlating Ti/Ca with other 
observables (e.g., metallicity, velocity-component structure) and by modeling the ionization 
properties of singly-ionized Ca and Ti in different environments.}
{We conclude that Ca\,{\sc ii} and Ti\,{\sc ii} bearing absorption-line systems trace 
predominantly neutral gas in the disks and inner halo regions of galaxies, where the 
abundance of Ca and Ti reflects the local metal and dust content of the gas.
Our study suggests that the Ti/Ca ratio represents a useful measure for the 
gas-to-dust ratio and overall metallicity in intervening absorption-line systems.}


\titlerunning{Dust depletion of Ca and Ti in QSO absorption line systems}
\authorrunning{Guber \& Richter}

\maketitle

%%%%%%%%%%%%%%%%%%%%%%%%%%%%%%%%%%%%%%%%%%%%%%%%%%%%%%%%%%%%%%%%%%%%%%%%%%%%%%%%%%%%%%%%%%%%%%%%%%%

\section{Introduction}

Systematic studies that are aiming at characterizing the chemical composition of gas and dust in the
interstellar medium (ISM) and circumgalactic medium (CGM) in and around galaxies are crucial for our
understanding of the formation and evolution of galaxies and their constituents.

Among the various observational strategies to measure the composition of the ISM the method of absorption
spectroscopy in the ultraviolet (UV) and in the optical regime against bright background point sources
stands out. It provides access to many diagnostic lines from low, intermediate, and high metal
ions that can be used to explore the properties of gas in the ISM in a wide range of physical properties.
In addition, from the observed deficiency of certain elements in the gas phase (commonly referred to
as ``depletion'') one can conclude on the composition and distribution of interstellar dust grains in
the ISM and their role in the complex astrochemical balance of star-forming sites in the Universe.
In our own Galaxy, UV and optical absorption-line spectroscopy can be carried out against a large number
of Galactic background stars near and far, providing a detailed insight into metal abundances of heavy
elements in the multi-phase ISM and dust-depletion patterns (for a review see \textbf{\citet{bjenkins2} and} \citet{bsavage}).
For external galaxies beyond the Local Group, however, absorption-line spectroscopy of their ISM is 
limited to sightlines against extragalactic background sources, such as all sorts of active galactic nuclei 
(AGN; throughout the following we will use the term ``QSO'' for all types of AGN that can be used
for absorption spectroscopy). 

QSO absorption-line systems (QALS) with the highest hydrogen column densities 
$(\log N($H\,{\sc i}$)>20.3)$ cause strongly saturated Lyman-alpha (Ly\,$\upalpha$) 
absorption in a QSO spectrum with prominent Lorentzian damping wings. Such systems
therefore are called Damped-Ly\,$\upalpha$ systems (DLAs). At low redshifts,
DLAs most likely are associated to neutral gas in galactic discs and the inner halo regions 
of galaxies; they contain most of the neutral gas mass in the Universe \citep{bprochaska}.
In sequence of further decreasing H\,{\sc i}-column densities QALS are called 
sub-damped Ly\,$\upalpha$-systems (Ly\,$\upalpha$ absorption line saturated but dominated 
by the Doppler-core), and Lyman-limit systems (LLS), because their $N($H\,{\sc i}$)$ is 
sufficient to produce continuous absorption for all absorber-restframe wavelengths 
smaller or equal the (redshifted) ionization edge of H\,{\sc i}. Sub-DLAs and LLS, which
contain an increasing fraction of ionized gas, are believed to characteristically 
arise in the extended gaseous halos of galaxies. 
Because QSOs are randomly distributed across the sky and because the cross section of the 
neutral gas regions in and around galaxies is small, QSO absorption-line studies of neutral
gas in galaxies at low $z$ beyond the Local Group require a large QSO data sample to detect 
DLAs, sub-DLAs, and LLS at a number that allows us to study their properties at a statistically 
significant level.

Next to the many lines of low, intermediate and high metal ions that can be observed in the FUV and
UV for low-redshift absorbers using space-based spectrographs (such as the Cosmic 
Origins Spectrograph (COS) on-board the \emph{Hubble Space Telescope} (HST) there are
a few absorption lines from metal ions in near-UV and in the blue part of the optical spectrum.
Of particular interest for absorption studies of neutral gas and dust in and around galaxies at 
low redshifts ($z\leq 0.5)$ are the transitions from singly-ionized calcium (Ca\,{\sc ii})
and singly-ionized titanium (Ti\,{\sc ii}).
Both, Ca\,{\sc ii} and Ti\,{\sc ii}, have near ultraviolet (UV) transitions (see Table\,\ref{Table I}), 
which are shifted into the visible part of the spectrum for low-redshift absorbers. 
This makes absorption from these lines detectable for powerful ground-based instruments 
(such as the Ultraviolet and Visual Echelle Spectrograph (UVES) installed at the 
 \emph{Very Large Telescope} (VLT)) that provide spectral data with a resolution and
signal-to-noise (S/N) that typically is substantially higher than that of space-based UV 
spectrographs.
Both elements, Ca and Ti, have comparatively high dust condensation temperatures 
of $T_\mathrm{c}\gtrsim \SI{1.5e+3}{\kelvin}$\footnote{$T_\mathrm{c}$ for a specific 
element is the temperature below which 50\,\% of the total 
amount is condensed into the dust phase (Wasson 1995).}. 
Thus, when interstellar gas cools down to temperatures below $T_\mathrm{c}$, these elements are 
expected to belong to the elements that condense into the dust phase first.
By coincidence, the ionization energy $E_\mathrm{ion}$ for Ti\,{\sc ii} nearly equals the one of 
neutral hydrogen. This means that both species live in the same interstellar gas phase.
The ionization energy for Ca\,{\sc ii} is slightly lower than for Ti\,{\sc ii} and H\,{\sc i}.
For systems containing warm neutral gas, the true gas amount of Ca in the neutral gas phase 
may be larger than the one inferred from Ca\,{\sc ii}.

Previous studies of Ti\,{\sc ii} and Ca\,{\sc ii} in the Milky Way, in the Small \& Large Magellanic Cloud,
and in intervening QALS at high redshift by \citet{bwelty} indicate that in low-metallicity systems
the Ti/Ca ratio is substantially raised compared to the Milky Way \citep{bwelty}.
This trend can be explained by the lower metal and dust content and different gas/dust properties 
in such systems. In this scenario, the increase of Ti/Ca ratio with {\it decreasing} 
metallicity/dust abundance then reflects the distinct dust depletion properties of the elements Ti 
and Ca: while Ca appears to be severely depleted even in low-metallicity gas, strong Ti depletion 
appears to be relevant only for high-metallicity environments (see \citet{bwelty}).
The dust-to-gas ratio and the ionizing radiation field (that strongly influences the abundance of 
Ca\,{\sc ii} in predominantly neutral gas) can, however, vary substantially within gaseous structures 
in and around a galaxies (see \citet{brichter}).
For instance, the dust-depletion patterns found in diffuse clouds in the Milky Way halo 
are remarkably distinct from those found in the dense regions in the Milky Way disk \textbf{with a more or less continuous transition between them} (\textbf{see \citet{bjenkins2} and} 
\citet{bsavage}). A systematic study of Ti\,{\sc ii} and Ca\,{\sc ii} in QALS at low and high 
redshift and their relation to galaxies therefore not only holds the prospect of increasing 
our understanding of dust abundances and depletion patterns in the interstellar media of 
gas-rich galaxies. It also provides important information on the distribution and properties of 
neutral gas structures beyond the stellar disks in galaxies, such as in high-velocity clouds and in 
circumgalactic tidal gas streams (e.g., Fox et al.\,2013,2014; Richter et al.\,2014). 

In this study, we examine 34 low-$z$ ($z\leq 0.5$) Ca\,{\sc ii}-selected QALS observed with 
VLT/UVES to study their Ti\,{\sc ii} abundance and the Ti\,{\sc ii}/Ca\,{\sc ii} ratios and
investigate the gas and dust properties of the absorbers.
We complement our data with literature information on Ti\,{\sc ii} and Ca\,{\sc ii} 
in the Milky Way, the Magellanic Clouds, and high-redshift absorbers.
We further collect information about the column densities of H\,{\sc i} and
Zn\,{\sc ii} in the absorbing systems, explore the association of 
absorbers with galaxies, study the ionization 
properties of Ti\,{\sc ii}/Ca\,{\sc ii} gas, and correlate the 
Ti\,{\sc ii}/Ca\,{\sc ii} ratios with other relevant quantities.

The paper is organized as follows. In Sect.\,2 we shortly describe the observations, 
the spectral data and the analyzing methods. In Sect.\,3 we discuss the absorption properties 
of the Ca\,{\sc ii}/Ti\,{\sc ii} systems and investigate important parameter correlations.
A comparison between the absorber properties in our QSO sample and other, local
Ca\,{\sc ii}/Ti\,{\sc ii} absorbing environments (Milky Way, Magellanic Clouds) 
is provided in Sect.\,4.
In Sect.\,5 we model the ionization conditions in Ca\,{\sc ii}/Ti\,{\sc ii} absorbers. 
Finally, a summary of our study is given in Sect.\,6.

%%%%%%%%%%%%%%%%%%%%%%%%%%%%%%%%%%%%%%%%%%%%%%%%%%%%%%%%%%%%%%%%%%%%%%%%%%%%%%%%%%%%%%%%%%%%%%%%%%%

\section{Observations, data, and analyzing methods}

\subsection{UVES data}

\begin{figure}
  \resizebox{\hsize}{!}{\includegraphics{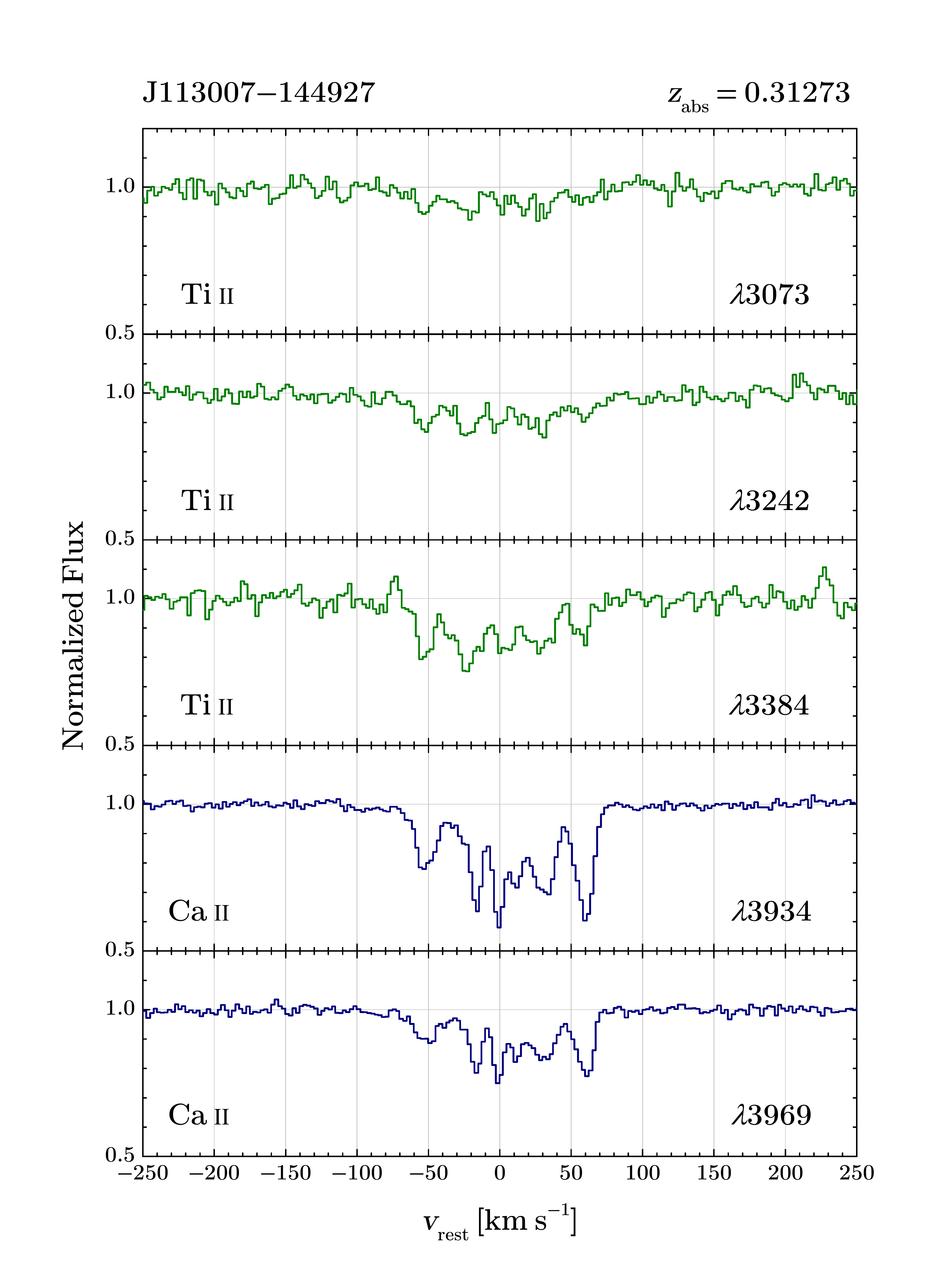}}
  \caption{
Example for velocity profiles of Ti\,{\sc ii} and Ca\,{\sc ii} in our    UVES spectra, here for the absorber at $z_\mathrm{abs}=0.31273$ towards J113007$-$14492. In this system, a multi-component structure is evident in both Ca\,{\sc ii} as well as in Ti\,{\sc ii}.}
  \label{samplespectrum2}
\end{figure}

For our study we examined 34 high-resolution\footnote{$R:=\lambda / \Delta \lambda \geq 45.000$} 
ESO UVES/VLT spectra of QSOs (with emission redshifts predominantly higher than $0.5$)
for intervening Ca\,{\sc ii} and Ti\,{\sc ii} absorption at redshifts $z \leq 0.5$. 
\textbf{A example spectrum is shown in Fig.\,\ref{samplespectrum2}.} Information about the examined QSO is provided in Table\,\ref{Table I}.

%%%%%%%%%%%%%%%%%%%%%%%%%%%%%%%%%%%%%%%%%%%%%%%%%%%%%%%%%%%%%%%%%%%%%%%%%%%%%%%%%%%%%%%%%%%%%%%%%%%

%%%%%%%%%%%%%%%%%%%%%%%%%%%%%%%%%%%%%%%%%%%%%%%%%%%%%%%%%%%%%%%%%%%%%%%%%%%%%%%%%%%%%%%%%%%%%%%%%%%

%%%%%%%%%%%%%%%%%%%%%%%%%%%%%%%%%%%%%%%%%%%%%%%%%%%%%%%%%%%%%%%%%%%%%%%%%%%%%%%%%%%%%%%%%%%%%%%%%%%

The study continues and extends our previous survey of intervening Ca\,{\sc ii} absorbers at low redshift presented in \citet{brichter}. The UVES 
data are publically available in the ESO archive. Here we use data that have been extracted and
pre-reduced as part of the SQUAD project (PI: M. Murphy) using a modified version of the
UVES reduction pipeline. For more details on the data reduction and the procedure of identifying 
intervening Ca\,{\sc ii} absorbers in the spectra see \citet{brichter}.

For the (independent) analysis of the absorbers we used the the ESO/MIDAS software executing 
the following steps.
First, we used the absorber redshifts given in \citet{brichter} to create restframe-velocity 
plots of all lines of interest (i.e., the lines from Ca\,{\sc ii} and Ti\,{\sc ii}, but also 
lines from Na\,{\sc i} and Mg\,{\sc ii}; see Table\,\ref{line data}).
We then rejected all blended absorption lines (i.e., lines that exhibit an unexpected absorption 
behavior according to the known different oscillator strengths of the transitions).
We re-normalized the spectra in the vicinity of every line of interest using low-order polynomials,
determined the local signal to noise ratio $(S/N)_\mathrm{vb}$ per velocity bin $\Delta v_\mathrm{pixel}$, 
and finally determined the observed equivalent width by a direct pixel integration. \textbf{Within the sample the $S/N$ per resolution element varies between 9 and 540.}
All the systems examined in this study show unsaturated Ca\,{\sc ii} and Ti\,{\sc ii} absorption in their
respective transitions. Therefore, we use the apparent-optical depth (AOD) method \citep{bsavage} to 
derive column densities for the various ions from the various absorption lines.
For each ion, we calculated the error-weighted mean column density from the values derived for each 
individual line
\begin{equation}
\overline{N}=\frac{\sum_\mathrm{i} p_\mathrm{i} N_\mathrm{i}}{\sum_\mathrm{i} p_\mathrm{i}};\,\,p_\mathrm{i}=\frac{1}{\Delta N_\mathrm{i}^2}.
\end{equation}
\textbf{We computed the uncertainty of the error-weighted mean value for the case of no relevant systematic errors (``internal consistency''),} 
\begin{equation}
\Delta \overline{N}=\sqrt{\frac{1}{\sum_\mathrm{i} p_\mathrm{i}}}
\end{equation}
\textbf{as well as the standard deviation of the weighted mean value (``external consistency'')}
\begin{equation}
\Delta \overline{N}= \sqrt{\frac{\sum_\mathrm{i}p_\mathrm{i}(N_\mathrm{i}-\overline{N})^2}{(n-1)\sum_\mathrm{i} p_\mathrm{i}}}
\end{equation}
\textbf{and rejected the smaller one of both error values.}
In case that at the wavelength of the (unblended) absorption line with the strongest oscillator strength 
no absorption feature is visible the determination of a $4\sigma$ upper limit (UL) for the column density is 
possible using the signal to noise ratio per resolution element (e.g., \citet{btumlinson}, \citet{brichter2}) via the relation

\begin{equation}
\label{N upper limit}
N_\mathrm{UL}=\SI{1.13e+20}{\per\centi\meter\squared} 
\frac{4}{R\, (S/N)_\mathrm{re}\, f \, (\lambda_0 /\SI{}{\angstrom})},
\end{equation}
where $R=\lambda/\Delta \lambda$ is the spectral resolution, $(S/N)_\mathrm{re}$ the signal-to noise ration per resolution element, $f$ the oscillator strength, and $\lambda_0$ the restframe wavelength of the transition.
\subsection{Supplementary data}

We supplement our own measurements with literature values of Ca\,{\sc ii} and/or Ti\,{\sc ii} column 
densities in 76 QALS at predominantly high ($z>1$) redshifts, as compiled by \citet{bwelty}. We also 
searched in the astrophysical database SIMBAD for galaxies within 2\,Mpc that could be associated with the 
absorbers and derived corresponding impact parameters. 

To relate the observed Ti/Ca ratios in intervening metal absorbers to the overall metallicity
of the absorbing gas, information on other observable metal ions are useful.
Among the various weakly-ionized metal ions, the element zinc (Zn) is particularly important
for this purpose.
Compared with Ti and Ca, the depletion of Zn under normal interstellar conditions
is expected to be very small and thus can be neglected (\citet{bsavage}; \citet{btimmes})\textbf{, although \citep{bjenkins2} found evidence for mild Zn-depletions in dense interstellar conditions}.
With an ionization energy larger than that of H\,{\sc i} singly-ionized zinc (Zn\,{\sc ii}) 
is assumed to be the dominant ionization state in neutral interstellar gas.
Therefore, the observed Zn\,{\sc ii}/H\,{\sc i} ratio serves as a robust measure for 
the absolute Zn abundance and the overall metallicity of the gas. For our study we therefore
have collected all available information on the Zn abundances in the sample of 
Ti/Ca absorption systems from the literature.

Throughout the paper we use a $\Lambda$CDM cosmology with $H_0=67.3\,\mathrm{km}\, \mathrm{s}^{-1} \, 
\mathrm{Mpc}^{-1}$, $\Omega_\mathrm{M}=0.315$ 
and $\Omega_\mathrm{V}=0.685$ \citep{bplanck}) and solar reference abundances from \citet{basplund}.
The results from our line fitting and literature search are presented in Tables A1$-$A4 in the
Appendix.

In this paper all column densities $N$ are given in \SI{}{\per\square\centi\metre}, all 3-dimensional
 particle densities in \SI{}{\per\cubic\centi\metre}.

%%%%%%%%%%%%%%%%%%%%%%%%%%%%%%%%%%%%%%%%%%%%%%%%%%%%%%%%%%%%%%%%%%%%%%%%%%%%%%%%%%%%%%%%%%%%%%%%%%%

\begin{table}
 \caption{Vacuum wavelengths and oscillator strengths used in this paper. 
  Values are taken from \citet{bmorton}.}
 \label{line data}
 \centering
 \begin{tabular}{lrr}
  \hline\hline
   Ion \& Transition     & $\lambda_\mathrm{vac}$       & $f$\\
                                                &$[\SI{}{\angstrom}]$   & \\
 \hline
Ti\,{\sc ii} $\lambda3073$ 	& 3073.863 	& 0.121\\
Ti\,{\sc ii} $\lambda3242$ 	& 3242.918 	& 0.232\\
Ti\,{\sc ii} $\lambda3384$ 	& 3384.730 	& 0.358\\
Ca\,{\sc ii} $\lambda3934$ 	& 3934.7750 & 0.6267\\
Ca\,{\sc ii} $\lambda3969$ 	& 3969.5901	& 0.3116\\
\hline
\end{tabular}
\end{table}

%%%%%%%%%%%%%%%%%%%%%%%%%%%%%%%%%%%%%%%%%%%%%%%%%%%%%%%%%%%%%%%%%%%%%%%%%%%%%%%%%%%%%%%%%%%%%%%%%%%

\begin{table*}
\centering
  \caption{Quasar data (obtained from SIMBAD) for sightlines with QALS for which we derived  Ti\,{\sc ii} column densities from VLT/UVES. Coordinates in ICRS(epJ2000).}
 \label{Table I}
 \centering
  \begin{tabular}{lllll}
  \hline\hline
Quasar name       & Alternative names     & $\alpha_{2000}$      & $\delta_{2000}$  & $z_\mathrm{em}$         \\
                  &                       & $[$h\,min\,s$]$      & $[\si{\degree} \, \si{\arcmin}\, \si{\arcsec}]$   &  \\
\hline
J000344$-$232355        & QSO J0003$-$2323, HE 0001$-$2340                      & 00 03 44.92   & $-$23 23 54.8        & 2.280\\
J005102$-$252846        & LBQS 0048-2545, QSO B0048$-$2545                      & 00 51 02.3    & $-$25 28 48          & 2.082\\
J012517$-$001828        & QSO B0122$-$005, PKS 0122$-$005                       & 01 25 17.1498 & $-$00 18 28.878      & 2.276\\
J014125$-$092843        & QSO J0141$-$0928, PKS 0139$-$09, QSO B1038$-$097 & 01 41 25.83179     &$-$09 28 43.6791      & 1.034\\
J030844$-$295702        & no quasar found in SIMBAD                             & $-$           & $-$                  & $-$\\
J042707$-$130253        & QSO J0427$-$1302, PKS 0424$-$131,QSO B0424$-$131      & 04 27 07.3003 & $-$13 02 53.644      & 2.166\\
J044117$-$431343        & QSO J0441$-$4313, QSO B0439$-$433,  & 04 41 17.32     & $-$43 13 45.4                        & 0.594\\
                                        & HE 0439$-$4319, PKS 0439$-$433        & & & \\
J045608$-$215909        & QSO J0456$-$2159, HE 0454$-$2203, PKS 0454$-$22       & 04 56 08.9272 & $-$21 59 09.543      & 0.534\\
J050112$-$015914        & 4C $-$02.19, QSO B0458$-$0203, PKS 0458$-$020         & 05 01 12.80988   & $-$01 59 14.2561  & 2.286\\
J051707$-$441055        & QSO B0515$-$4414, 4E 0515$-$4414                      & 05 17 07.63   & $-$44 10 55.5        & 1.713\\
J055158$-$211949        & QSO B0549$-$213, QSO J0551$-$2199                     & 05 51 58.3    & $-$21 19 50          & 2.245\\
J094253$-$110426        & QSO B0940$-$1050, QSO J0942$-$1104, HE 0940$-$1050    & 09 42 53.49   & $-$11 04 25.9        & 3.054\\
J095456$+$174331        & QSO B0952$+$179, PKS 0952$+$179                       & 09 54 56.82362   & $+$17 43 31.222   & 1.475\\
J102837$-$010027        & LBQS 1026$-$0045B, QSO B1026$-$0045B                  & 10 28 37.0163 & $-$01 00 27.555      & 1.532\\
J104733$+$052454        & QSO B1044$+$056                                       & 10 47 33.1551 & $+$05 24 54.882      & 1.334\\
J110325$-$264515        & QSO B1101$-$26, PG 1101$-$264                         & 11 03 25.31   & $-$26 45 15.875      & 2.145\\
J113007$-$144927        & QSO J1130$-$1449, PKS 1127$-$145, QSO B1127-14        & 11 30 07.0473 & $-$14 49 27.424      & 1.189\\
J121140$+$103002        & LBQS 1209$+$1046, QSO B1209$+$1046                    & 12 11 40.5903 & $+$10 30 02.026      & 2.193\\
J121509$+$330955        & QSO J1215$+$3309, Ton 1480, QSO B1212$+$3326          & 12 15 09.2208 & $+$33 09 55.233      & 0.615\\
J123200$-$022404        & 4C $-$02.55, QSO B1229$-$021, PKS 1229$-$021          & 12 32 00.160  & $-$02 24 04.794      & 1.043\\
J124646$-$254749        & QSO J1246$-$2547, PKS 1244$-$255,     QSO B1244$-$255 & 12 46 46.8006 & $-$25 47 49.371      & 0.638\\
J133007$-$205616        & QSO B1327$-$2040, PKS 1327$-$206                      & 13 30 07.7    & $-$22 56 16.5        & 1.165\\
J142249$-$272756        & QSO B1420$-$272, PKS 1419$-$272                       & 14 22 49.2317 & $-$27 27 56.857      & 0.985\\
J142253$-$000149        & QSO J1422$-$001, QSO B1420$-$0011                     & 14 22 53.3157 & $-$00 01 49.077      & 1.083\\
J144653$+$011356        & LBQS 1444$+$0126, QSO B1444$+$0126                    & 14 46 53.0    & $+$01 13 55          & 2.210\\
J162439$+$234512        & 4C 23.43, QSO B1622$+$238, PKS 1622$+$238             & 16 24 39.0847 & $+$23 45 12.183      & 0.927\\
J215501$+$092224        & QSO J2155$-$0922, PHL 1811, QSO B2152$-$0936          & 21 55 01.5152 & $+$09 22 24.688      & 0.192\\
J220743$-$534633        & QSO B2204$-$54, PKS 2204$-$540                        & 22 07 43.7420 & $-$53 46 33.826      & 1.206\\
J224752$-$123719        & QSO B2245$-$1282, PKS 2245$-$128                      & 22 47 52.6411 & $-$12 37 19.721      & 1.892\\
J231359$-$370446        & QSO B2311$-$373, PKS 2311$-$373                       & 23 13 59.7    & $-$37 04 46          & 2.476\\
J232820$+$002238        & QSO J2328$+$0022, QSO B2325$+$0006                    & 23 28 20.3797 & $+$00 22 38.262      & 1.302\\
J235731$-$112539        & QSO B2354$-$117, PKS 2354$-$117                       & 23 57 31.1976 & $-$11 25 39.176      & 0.960\\
\hline
\end{tabular}
\end{table*}

%%%%%%%%%%%%%%%%%%%%%%%%%%%%%%%%%%%%%%%%%%%%%%%%%%%%%%%%%%%%%%%%%%%%%%%%%%%%%%%%%%%%%%%%%%%%%%%%%%%

\section{Absorber properties and classification}

To characterize the absorption properties of the systems in our sample we start our analysis by 
looking for correlations between the measured column densities of Ca\,{\sc ii}, Ti\,{\sc ii}, H\,{\sc i},
and Zn\,\textsc{ii}. Our analysis strategy is summarized as follows.

First, we divide our absorbers in different sub-samples, depending on whether Ti\,\textsc{ii} is 
detected or not. We then explore Ti/Ca-ratios in our absorption systems and compare them with previous 
measurements in the Milky Way and in the Magellanic Clouds (Sect.\,3.1). 
For systems for which H\,{\sc i} information is available we investigate the relation between 
Ca\,{\sc ii}, Ti\,{\sc ii} and the total neutral gas column density (Sect.\,3.2).
In Sect.\,3.3 we explore the number density of intervening Ti\,\textsc{ii}-absorbers. 

From our 34 absorption-line systems, 27 exhibit absorption in Ca\,{\sc ii}, but only 5 show 
significant absorption in Ti\,{\sc ii}. For all systems we have measured equivalent widths 
(equivalent-width limits) and column densities (column-density limits) of Ca\,\textsc{ii} 
and/or Ti\,\textsc{ii} following the procedure outlined in Sect.\,2.
Because of the inhomogeneity of the absorber sample we divide our systems into different samples
according to the detection/non-detection of the different metal ions:

\begin{enumerate}
\item Sample A: Ca\,{\sc ii} systems with known $N$(Ti\,{\sc ii}),
\item Sample B: Ca\,{\sc ii} systems with upper limits for $N$(Ti\,{\sc ii}),
%\item Sample C: Absorption systems with known column densities or upper 
%                limits for Ti\,{\sc ii} but without any Ca\,{\sc ii} information, and
%\item Sample D: Systems with only upper limits both for $N$(Ca\,{\sc ii}) and $N$(Ti\,{\sc ii}) 
%                and systems without any information about $N($Ti\,{\sc ii}$)$.
\end{enumerate}

Information about Samples A \& B are compiled in Table\,\ref{tab class 1 and 2} (Appendix). 
This table also contains all Ca\,\textsc{ii} systems taken from \citet{bwelty} that can be classified as sample A or B.
Table \,\ref{tab clas 3} in the Appendix gives information about all Ti\,\textsc{ii} systems 
from \citet{bwelty} without any information about Ca\,\textsc{ii}-column densities.
Finally, Table \,\ref{tab data cl 4} (Appendix) lists all the systems of our data sample 
and from \citet{bwelty} with only upper limits for the column densities of both, 
Ca\,\textsc{ii} and Ti\,\textsc{ii} and Ca\,\textsc{ii} systems without any information 
about Ti\,\textsc{ii}. 

The absorbers considered in this study span a huge
redshift range between $z_{\rm abs}=0.003$ and $3.774$. The redshift-dependence of the
observed absorber properties will be discussed in Sect.\,4.

%%%%%%%%%%%%%%%%%%%%%%%%%%%%%%%%%%%%%%%%%%%%%%%%%%%%%%%%%%%%%%%%%%%%%%%%%%%%%%%%%%%%%%%%%%%%%%%%%%%

\begin{figure}
  \resizebox{\hsize}{!}{\includegraphics{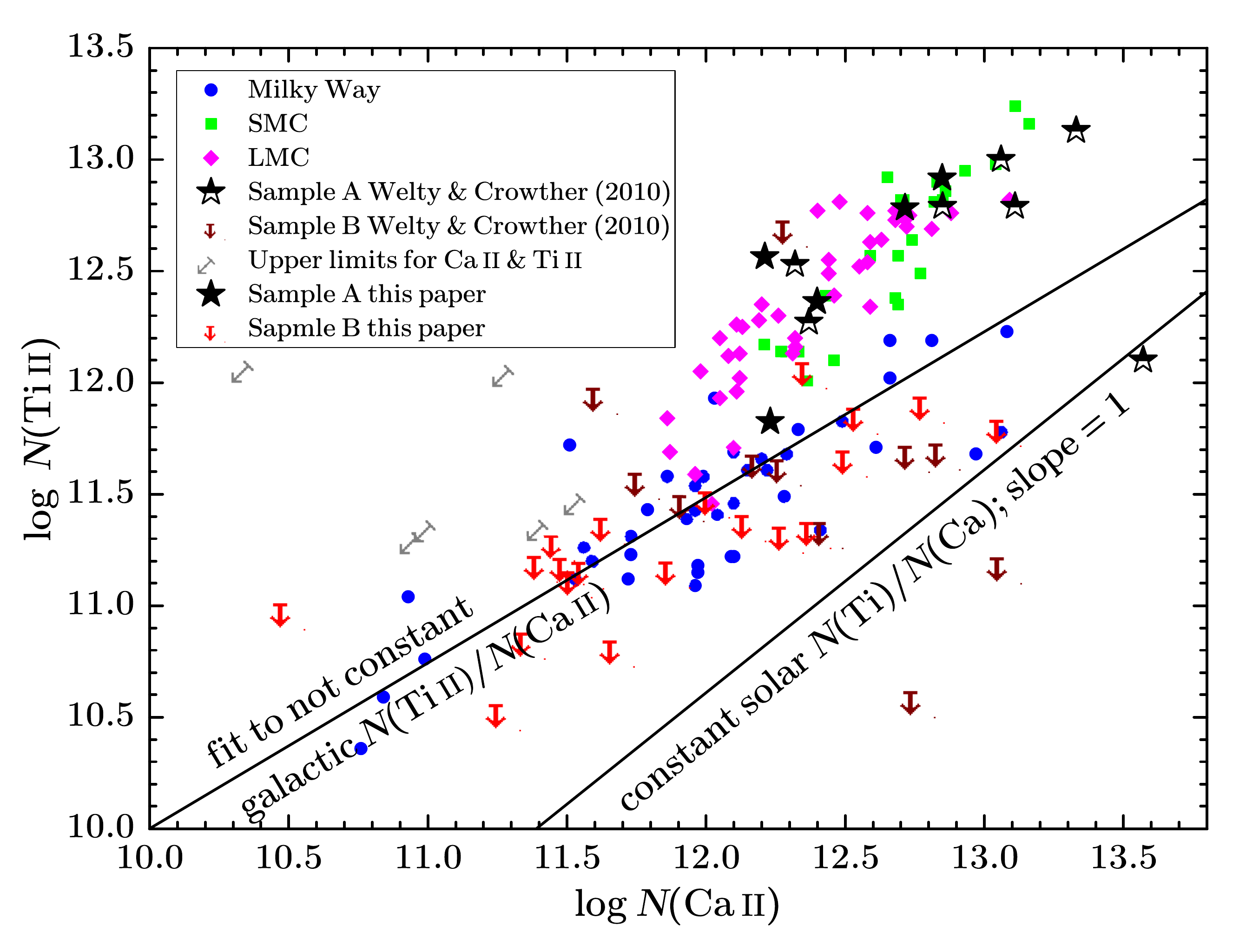}}
  \caption{$\log [N($Ti\,{\sc ii}$)]$ vs. $\log [N($Ca\,{\sc ii}$)]$ for QALS for which Ca\,{\sc ii} 
 and Ti\,{\sc ii} column densities (sample A) or upper limits therefore (sample B) are available. 
 ALS in the Magellanic Clouds as well as in the Milky Way are shown, too. With the one exception 
 showing a solar like Ti/Ca-ratio all the sample A systems have supersolar Ti/Ca-ratios. Systems 
 with constant, solar Ti/Ca-ratio would lay on the black line with slope 1. The other black line 
 represents the linear fit of the Galactic ALS already published by \citet{bwelty}. Its slope $<$ 1 
 indicates decreasing Ti/Ca-ratio with increasing Ca\,{\sc ii}-column density. }
  \label{overview}
\end{figure}

%%%%%%%%%%%%%%%%%%%%%%%%%%%%%%%%%%%%%%%%%%%%%%%%%%%%%%%%%%%%%%%%%%%%%%%%%%%%%%%%%%%%%%%%%%%%%%%%%%%

%%%%%%%%%%%%%%%%%%%%%%%%%%%%%%%%%%%%%%%%%%%%%%%%%%%%%%%%%%%%%%%%%%%%%%%%%%%%%%%%%%%%%%%%%%%%%%%%%%%

\subsection{Observed Ti/Ca ratios}

In Fig.\,\ref{overview} we plot $\log N($Ti\,{\sc ii}$)$ vs. $\log N($Ca\,{\sc ii}$)$ for the absorbers in
samples A and B.

For comparison, the Magellanic Cloud absorbers and high-redshift systems presented by \citet{bwelty}
are also shown. In addition, we indicate the solar Ti/Ca ratio and the measured trend for 
$\log N($Ti\,{\sc ii}$)$ vs. $\log N($Ca\,{\sc ii}$)$ in the Milky Way ISM with black solid lines.

For the Ca\,{\sc ii} systems with known values for $N($Ti\,{\sc ii}$)$ (sample A) we find that
all except one of the systems have Ti\,{\sc ii}/Ca\,{\sc ii} ratios that lie clearly above 
the solar Ti/Ca ratio and also above the relation observed in the Milky Way ISM.
Comparison with the data points from Welty \& Crowther (2010; data points) instead indicates 
that the systems from sample A are distributed in the same region in this plot where also the 
SMC and LMC absorbers are found.

For the Ca\,{\sc ii} systems without known Ti\,{\sc ii} column densities (sample B) the 
interpretation of the observed distribution of data points (=limits) is less straight-forward.
All the examined systems were studied with high resolution spectra that have (in most cases) 
comparable signal-to-noise ratios. Therefore, one can expect to obtain similar upper limits 
for $N$(Ti\,{\sc ii}) for most of the absorbers, independent of $N($Ca\,{\sc ii}$)$.
This indeed is the case: $75\,\%$ of the class 2 systems have upper limits (ul) of 
$\log N($Ti\,{\sc ii}$)_\mathrm{ul}\in[11-12]$.
Systems from sample B that have $\log N($Ca\,{\sc ii}$)<12$ do not provide useful information 
because their values for $N($Ti\,{\sc ii}$)_\mathrm{ul}$ are so high, that they could have Ti/Ca 
ratios that are either lower, equal or larger than the solar reference value.

In contrast, for the systems in sample B that have $\log N($Ca\,{\sc ii}$)>12$ (especially 
the ones with $\log N($Ca\,{\sc ii}$)>12.5$) we can conclude that they exhibit Ti/Ca ratios 
that are comparable to or even lower than the solar one (see Fig.\,\ref{overview}).
For the systems with $\log N($Ca\,{\sc ii}$)\ga 12$, the lack of absorbers showing Ti/Ca 
ratios between solar and supersolar solar values with $\left[ \mathrm{Ti}/\mathrm{Ca}\right]\ga 1$ 
indicates that Ca\,{\sc ii} absorbers can be subdivided into two classes:

\begin{enumerate}
\item Class 1: systems with $\left[ \mathrm{Ti}/\mathrm{Ca}\right]\ga 1$, and
\item \textbf{Class 2: systems with $\left[ \mathrm{Ti}/\mathrm{Ca}\right] \lesssim 0$.}
\end{enumerate}

%%%%%%%%%%%%%%%%%%%%%%%%%%%%%%%%%%%%%%%%%%%%%%%%%%%%%%%%%%%%%%%%%%%%%%%%%%%%%%%%%%%%%%%%%%%%%%%%%%%

\begin{figure}
  \resizebox{\hsize}{!}{\includegraphics{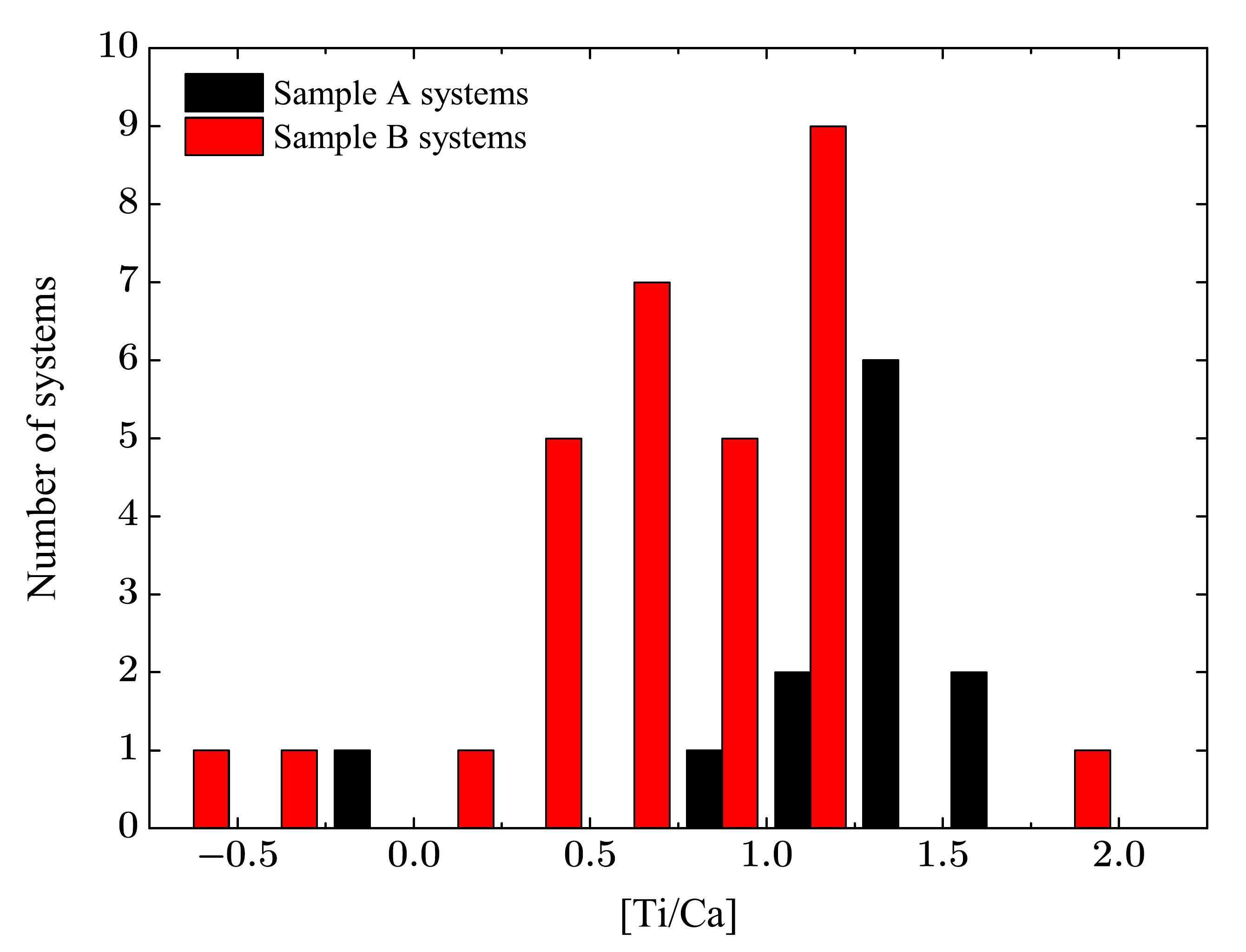}}
  \caption{Histogram comparing the Ti/Ca-ratios of sample A \& B systems. Except the one sample A 
  system with $[\mathrm{Ti}/\mathrm{Ca}]=-0.08$ all sample A systems are high Ti/Ca systems with 
  $[\mathrm{Ti}/\mathrm{Ca}]\gtrsim 1$.}
  \label{histogram}
\end{figure}

%%%%%%%%%%%%%%%%%%%%%%%%%%%%%%%%%%%%%%%%%%%%%%%%%%%%%%%%%%%%%%%%%%%%%%%%%%%%%%%%%%%%%%%%%%%%%%%%%%%

In Fig.\,\ref{histogram} we show a histogram for absorbers from samples A and B that displays 
the number of corresponding systems in specific intervals for 
$\left[ \mathrm{Ti}/\mathrm{Ca}\right]$.
Except the one system from sample A that has $\left[ \mathrm{Ti}/\mathrm{Ca}\right]=-0.08$, 
all sample A absorbers have $\left[ \mathrm{Ti}/\mathrm{Ca}\right]$ ratios according to the 
above defined class 1 ($\left[ \mathrm{Ti}/\mathrm{Ca}\right]\ga 1$). Sample-A absorbers have 
a mean $\left[ \mathrm{Ti}/\mathrm{Ca}\right]$ ratio of 
$\langle\left[ \mathrm{Ti}/\mathrm{Ca}\right]\rangle=1.34$, where the uncertainty for the 
individual measurement is 0.21 dex and an uncertainty for the average value is 0.07 dex. 
\textbf{Note that the individual values for sample B systems are only upper limits, so that the true
mean value of this group is even lower than the mean value of upper limits $\langle \left[ \mathrm{Ti}/\mathrm{Ca}\right]_\mathrm{ul} \rangle=0.8$.
Although within sample B there might be a few systems with high Ti/Ca ratios belonging 
to Class 1, most of them exhibit Ti/Ca-ratio limits that are more comparable to or even lower than the solar value ($\rightarrow$ class 2).} 
Except the one outlier sample A system with $\left[ \mathrm{Ti}/\mathrm{Ca}\right]\approx 0$, 
there is a one-to-one correspondence between sample A absorbers and class 1 systems.

In the following sections, we will further investigate whether the separation into two
different absorber classes is a result of the different dust (depletion) properties in the gas,
or is caused by ionization effects, or by a combination of these two aspects.

\subsection{Associated H\,{\sc i} absorption}

Only for a small sub-sample of the Ca\,{\sc ii}/Ti\,{\sc ii} systems in our sample literature 
values for the H\,{\sc i} column densities (from their Ly\,$\alpha$ absorption and/or \SI{21}{\centi\metre} emission) are available 
(see Tables\, \ref{tab class 1 and 2} \& \ref{tab data cl 4} for corresponding references).

In Fig.\,\ref{TiCavsH} we show the H\,{\sc i} column density dependence for 
$N($Ca\,{\sc ii}$)$ (left panel) and $N($Ti\,{\sc ii}$)$ (right panel) for absorbers in 
samples A and B.
There are a number of interesting trends visible that can be summarized as follows:

%%%%%%%%%%%%%%%%%%%%%%%%%%%%%%%%%%%%%%%%%%%%%%%%%%%%%%%%%%%%%%%%%%%%%%%%%%%%%%%%%%%%%%%%%%%%%%%%%%%

\begin{figure}
  \resizebox{\hsize}{!}{\includegraphics{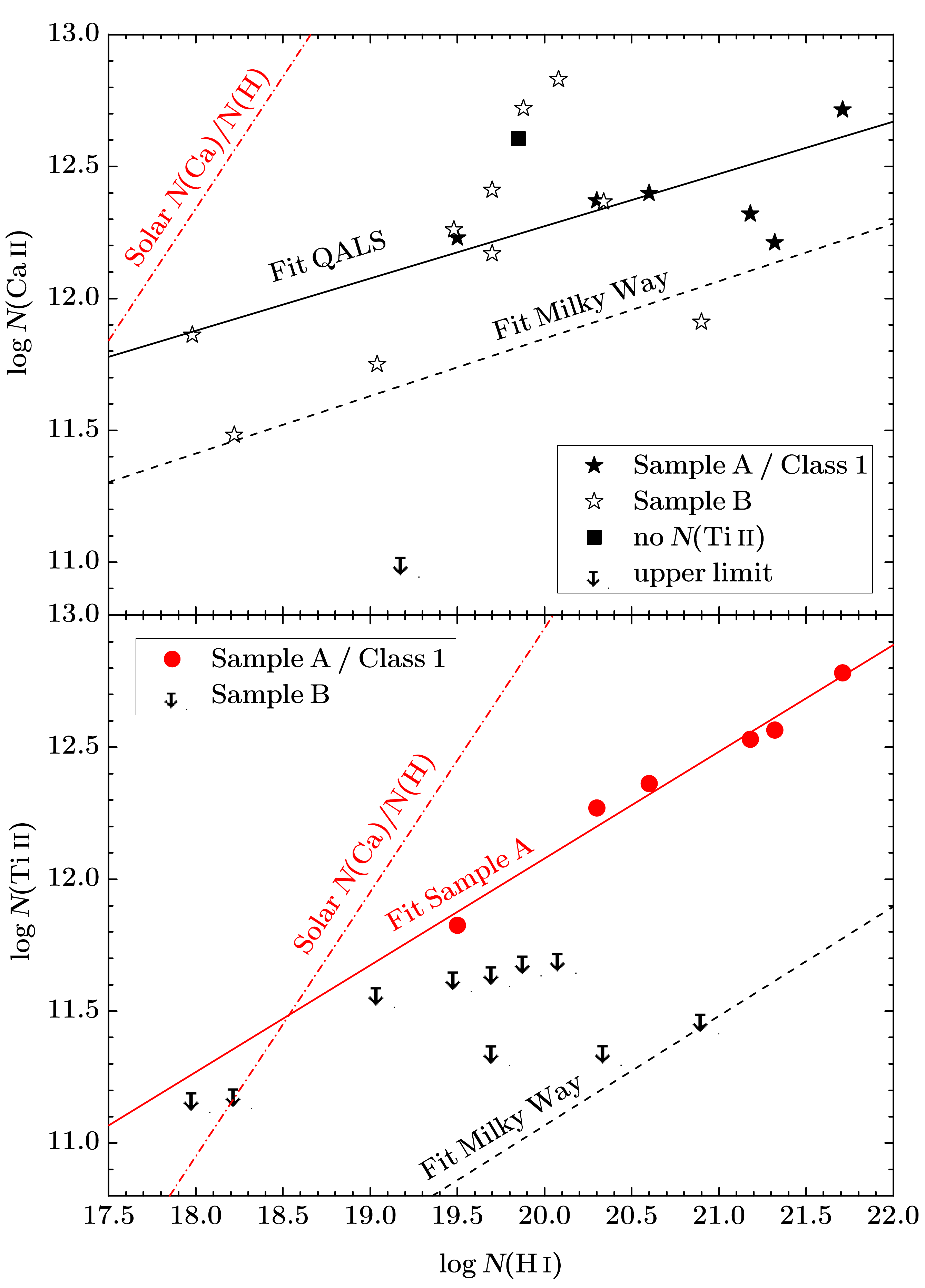}}
  \caption{$\log N(\mathrm{Ca}\,\textsc{ii})$ and $\log N(\mathrm{Ti}\,\textsc{ii})$ 
  vs. $\log N(\mathrm{H}\,\textsc{i})$.
  Every sample A system in the plots above is also a class 1 system.  
  Filled box in the Ca\,\textsc{ii}-plot: the one Ca\,\textsc{ii} system without any 
  Ti\,\textsc{ii}-information; the arrow represents the one Ca\,\textsc{ii} system with 
  only upper limits both for Ca\,\textsc{ii} \& Ti\,\textsc{ii}. 
  In case Ca\,\textsc{ii} or Ti\,\textsc{ii} is the dominant ionization state, systems 
  with solar Ca/H (Ti/H) ratio would lay on the line-dotted lines with slope 1.
  The dashed lines represent linear fits of galactic values (Ca\,\textsc{ii} from \citet{bwakker}, 
  Ti\,\textsc{ii} from \citet{bwelty}).}
  \label{TiCavsH}
\end{figure}

%%%%%%%%%%%%%%%%%%%%%%%%%%%%%%%%%%%%%%%%%%%%%%%%%%%%%%%%%%%%%%%%%%%%%%%%%%%%%%%%%%%%%%%%%%%%%%%%%%%

\begin{enumerate}

\item For both, Ca\,{\sc ii} and Ti\,{\sc ii} systems from sample A the column densities 
increase with increasing $N($H\,{\sc i}$)$. The relation for Ti\,{\sc ii} is relatively tight,
while for Ca\,{\sc ii} there is considerable scatter. From a fit of the data points
we derive a slope of $A=0.198 \pm 0.077$ (offset $n=8.3 \pm 1.6$) for the relation
log $N($Ca\,{\sc ii}$)=A\,$log $N($H\,{\sc i}$)+n$ and 
$B=0.404 \pm 0.031$ (offset $m=4.00 \pm 0.65$) for the relation
log $N($Ti\,{\sc ii}$)=B\,$log $N($H\,{\sc i}$)+m$.

\item For a given $N($H\,{\sc i}$)$ the column densities of Ca\,{\sc ii} and Ti\,{\sc ii} 
(sample A) are substantially larger than typically observed in the Milky Way.
The corresponding offsets are $\sim+0.4$ dex for Ca\,{\sc ii} and $\sim+0.9$ dex for 
Ti\,{\sc ii} (see \citet{brichter} \& \citet{bwelty}).

\item The slopes of the above given fits for both QALS and Milky Way clouds are much smaller 
than unity. This trend indicates decreasing Ca/H and Ti/H ratios in the gas phase for 
increasing $N($H\,{\sc i}$)$, a typical signature of column-density dependent dust depletion.
In addition, the different slopes of Ca\,{\sc ii} and Ti\,{\sc ii} suggest that Ti/Ca in the 
gas phase is increasing for increasing $N($H\,{\sc i}$)$, a signature for non-uniform 
differential depletion properties of these two elements; this aspect will be further 
investigated below.

\item The Ti\,{\sc ii} upper limits of the systems in sample B (presumably class 2 absorbers) are 
in contradiction to the Ti\,{\sc ii} column densities in sample A (class 1; Fig.\,\ref{TiCavsH}). All the absorbers in sample B have $\log [N($Ti\,{\sc ii}$)]<11.8$, even for 
comparatively high $N($H\,{\sc i}$)$. This trend supports our previously outlined scenario 
in which the absorbers can be separated into two sub-classes:
absorbers with comparatively high Ti/Ca ratios $([\mathrm{Ti}/\mathrm{Ca}]\ga 1)$ (class 1), 
that also have high Ti\,{\sc ii}/H\,{\sc i} ratios, and absorbers with low Ti/Ca ratios 
$([\mathrm{Ti}/\mathrm{Ca}] \approx0)$ (class 2), that also have low Ti\,{\sc ii}/H\,{\sc i} 
ratios.

\end{enumerate}

%%%%%%%%%%%%%%%%%%%%%%%%%%%%%%%%%%%%%%%%%%%%%%%%%%%%%%%%%%%%%%%%%%%%%%%%%%%%%%%%%%%%%%%%%%%%%%%%%%%

\begin{figure}
  \resizebox{\hsize}{!}{\includegraphics{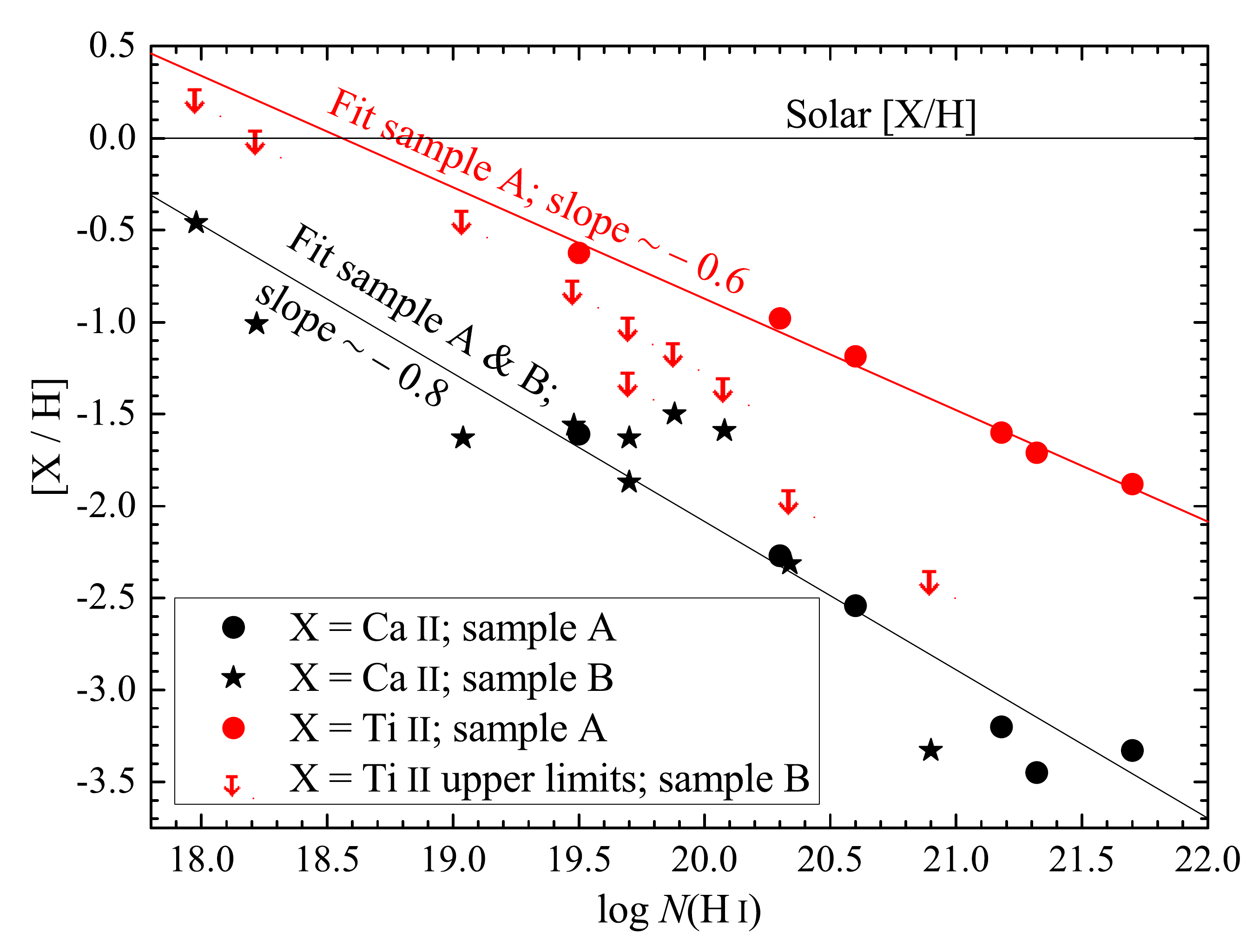}}
  \caption{$[$Ca$/$H$]$ and $[$Ti$/$H$]$ vs. $\log N(\mathrm{H\,\textsc{i}})$. In case of 
  Ti\,\textsc{ii} the solid line represents the linear fit of the sample A systems, which 
  in this case are all class 1 systems, too. In case of Ca\,\textsc{ii} the solid line 
  represents the linear fit of all systems with known $N(\mathrm{Ca\,\textsc{ii}})$. 
  Systems with solar Ca/H or Ti/H ratios would lay on the horizontal 
  solid line with $[\mathrm{X}/\mathrm{H}]=0$.}
  \label{TiHCaHvsH}
\end{figure}

%%%%%%%%%%%%%%%%%%%%%%%%%%%%%%%%%%%%%%%%%%%%%%%%%%%%%%%%%%%%%%%%%%%%%%%%%%%%%%%%%%%%%%%%%%%%%%%%%%%

As mentioned above, the observed trends for Ca\,{\sc ii}/H\,{\sc i} and Ti\,{\sc ii}/H\,{\sc i} 
in our survey indicate non-uniform differential dust depletions for Ca and Ti in the absorbing 
gas environments. To further investigate this aspect in Fig.\,\ref{TiHCaHvsH} we show the 
ratios $[\mathrm{Ca}/\mathrm{H}]$ or $[\mathrm{Ti}/\mathrm{H}]$ as function of 
log $N($H\,{\sc i}$)$.
Both the $N($H\,{\sc i}$)$-dependency of $[\mathrm{Ca}/\mathrm{H}]$ (for all the Ca\,{\sc ii} 
absorbers) and $[\mathrm{Ti}/\mathrm{H}]$ (only for sample A/class 1) can well be described by 
(linear) relations. The scatter around the fitted 
relations is more severe for $[\mathrm{Ca}/\mathrm{H}]$ than for $[\mathrm{Ti}/\mathrm{H}]$ 
(class 1), as expected from the trend seen in Fig.\,\ref{TiCavsH} 
(see also \citet{brichter}, their Fig.\,9).
In addition, the slope for $[\mathrm{Ca}/\mathrm{H}]$ ($-0.806 \pm 0.075$) is slightly 
steeper than the one for $[\mathrm{Ti}/\mathrm{H}]$ ($-0.599 \pm 0.032$).
For the sample A/class 1 systems this again indicates that the Ti/Ca ratios {\it increase} with 
increasing $N($H\,{\sc i}$)$.

%%%%%%%%%%%%%%%%%%%%%%%%%%%%%%%%%%%%%%%%%%%%%%%%%%%%%%%%%%%%%%%%%%%%%%%%%%%%%%%%%%%%%%%%%%%%%%%%%%%

\begin{figure}
  \resizebox{\hsize}{!}{\includegraphics{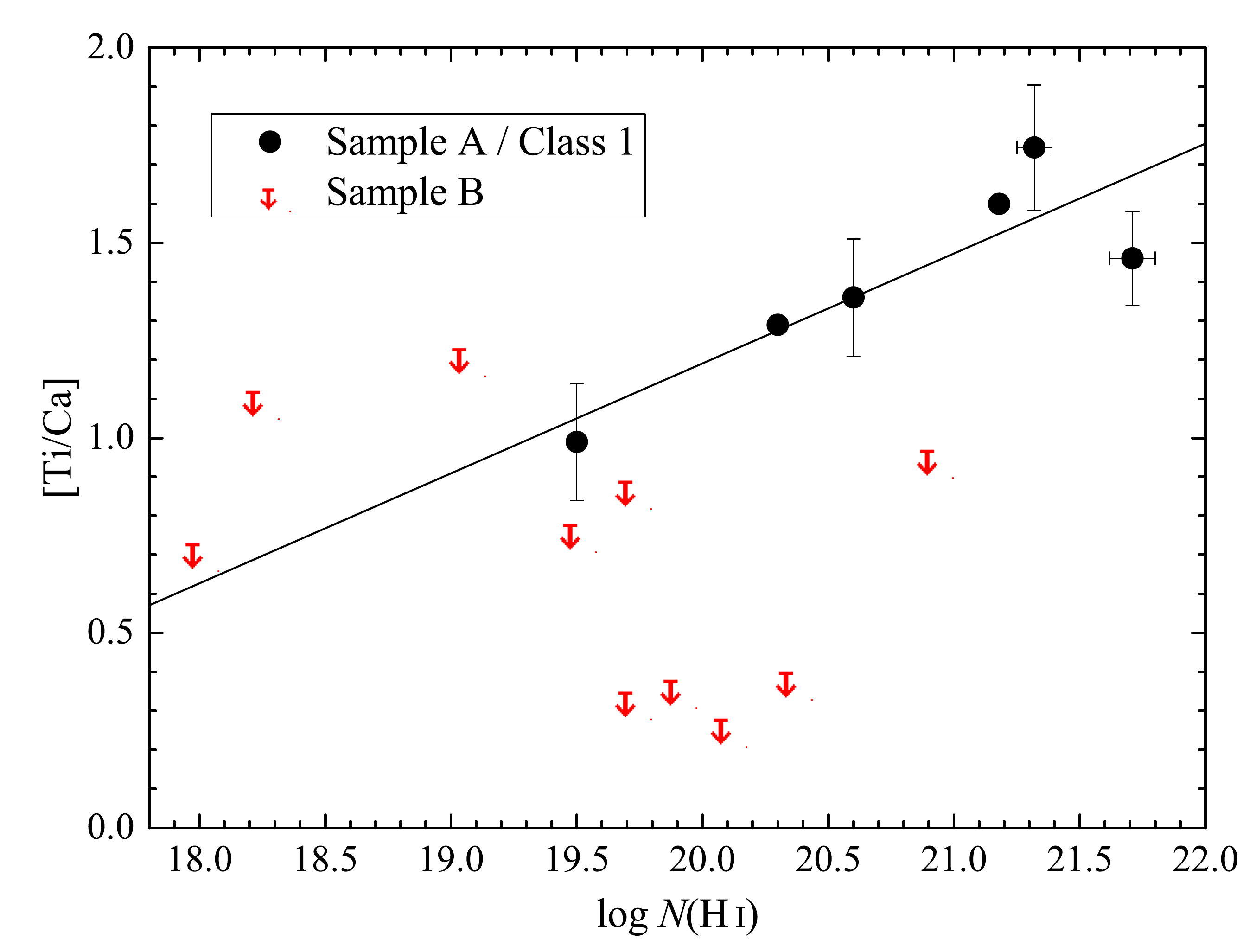}}
  \caption{$[\mathrm{Ti}/\mathrm{Ca}]$ vs. $\log N(\mathrm{H\,\textsc{i}})$ for sample A \& B systems. 
  All sample A systems in this plot are also class 1 systems. The black line represents the non weighted 
  linear fit of the sample A data. The distribution of the class 2 systems can't be described by this fit.}
  \label{TiCavslogH}
\end{figure}

%%%%%%%%%%%%%%%%%%%%%%%%%%%%%%%%%%%%%%%%%%%%%%%%%%%%%%%%%%%%%%%%%%%%%%%%%%%%%%%%%%%%%%%%%%%%%%%%%%%

If we plot $[$Ti$/$Ca$]$ directly against log $N($H\,{\sc i}) (Fig.\,\ref{TiCavslogH}), 
we see this trend directly for sample A/class 1 systems, where the [Ti/Ca]-H\,{\sc i} relation 
can be well fitted linearly with a slope of $+0.282 \pm 0.083$.
\textbf{Systems in sample B show a different behavior than sample A/class 1 systems and do not all lay on or above the trend for class 1 systems.}

\subsection{Number density of intervening Ti\,{\sc ii} absorbers}

Only five out of the 34 originally selected Ca\,{\sc ii} absorbers at $z<0.5$ from
the sample of 270 (304) QSOs presented in \citep{brichter} show associated Ti\,{\sc ii} absorption 
(see Table\, \ref{tab class 1 and 2}), suggesting that intervening Ti\,{\sc ii} absorbers in the local
Universe are very rare. This is, however, not surprising at all in view
of the substantially lower cosmic abundance of Ti compared to Ca (log $(\mathrm{Ti}/\mathrm
{H})_{\sun}=-7.05$ vs.
log $(\mathrm{Ca}/\mathrm{H})_{\sun}=-5.66$; \citet{basplund}) and the relatively small oscillator 
strengths of the available Ti\,{\sc ii} transitions compared to the Ca\,{\sc ii}
H\&K lines (Table\,\ref{line data}). Thus, for a typical detection limit of log $N$(Ti\,{\sc ii}$)=12.0$
in UVES data and a typical metallicity of $\leq 0.1$ solar, one requires a neutral
gas column as large as log $N$(H\,{\sc i}$)>20$, even if the dust depletion of Ti is zero.
Ti\,{\sc ii}-absorbing systems that do not show significant 
Ca\,{\sc ii} absorption are not expected to exist (except Ca\,{\sc ii} is blended by 
other lines or is located outside the observable wavelength range).
The detection of Ti\,{\sc ii} in an intervening absorber thus depends on three parameters: 
the total gas column density, the overall metallicity, and the depletion of Ti into dust grains.

All five Ti\,{\sc ii} absorbers in the UVES data sample analyzed in this study represent 
class 1 absorbers with high Ti\,{\sc ii}/Ca\,{\sc ii} ratios (Table\, \ref{tab class 1 and 2}) spanning 
a column density range of log $N$(Ti\,{\sc ii}$)=11.8-13.0$. Based on the 
distribution of $S/N$ in UVES data set (described in detail by \citet{brichter}) and 
using equation 1 together with at atomic data of the available Ti\,{\sc ii} transitions 
(Table 1) we calculate that all the 270 selected UVES spectra from the Ca\,{\sc ii} survey 
of \citep{brichter} are sensitive to Ti\,{\sc ii} column densities of 
log $N$(Ti\,{\sc ii}$)\geq 12.0$, where the total redshift path to detect Ca\,{\sc ii}/Ti\,{\sc ii} 
systems in these spectra is $\Delta z=89.15$. Four out of the five 
Ti\,{\sc ii} absorbers have Ti\,{\sc ii} column densities above this limit, so that
we derive a number density of intervening Ti\,{\sc ii} absorbers with 
log $N$(Ti\,{\sc ii}$)\geq 12.0$ at $z<0.5$ of $\mathrm{d}{\cal N}/\mathrm{d}z=4/270=0.045\pm 0.023$.

Because of a potential selection bias in the UVES absorber sample 
(see the detailed discussion in \citet{brichter}) this value might
be overestimated by up to 40 per cent, so that the true number density might 
be even lower. If we correct for this bias and reduce the
number density by 40 per cent we obtain $\mathrm{d}{\cal N}/\mathrm{d}z=0.027\pm 0.017$.

This value now can be compared to the number density of sub-DLAs and DLAs
absorbers at low redshift. Three of the low-redshift absorbers in our UVES sample 
are high-column density systems with log $N$(H\,{\sc i}$)>20.1$
(while for the fourth system no information on log $N$(H\,{\sc i}) is available).
All but one of the high-redshift Ti\,{\sc ii} absorbers with log $N$(Ti\,{\sc ii}$)\geq 12.0$ 
compiled by \citet{bwelty} have log $N$(H\,{\sc i}$)>20.1$, suggesting that 
Ti\,{\sc ii} absorption at this column-density level typically arises in 
DLAs and high-column density sub-DLAs (see Table A2 in the Appendix).

From the \SI{21}{\centi\metre} H\,{\sc i} mass function in the local Universe
Zwaan et al.\,(2005) derives $\mathrm{d}{\cal N}/\mathrm{d}z=0.060$ 
for absorbers with log $N$(H\,{\sc i}$)>20.1$, a number density that
is $2.2$ times higher than the estimated $\mathrm{d}{\cal N}/\mathrm{d}z$(Ti\,{\sc ii}$)=0.027$ 
at low redshift. From this we conclude that only $\sim 45$ per cent of 
high-column density sub-DLAs and DLAs exhibit Ti\,{\sc ii} absorption 
with log $N$(Ti\,{\sc ii}$)\geq 12.0$.
On the other hand, for more than half ($\sim 55$ per cent) of such
absorbers the combined effect of low metallicity and dust depletion
leads to a reduction of Ti in the gas phase in these systems by 
at least 1 dex when compared to the solar Ti reference abundance 
($20.10-12.00-7.05=1.05$; see above).

In Sect.\,4.2.2 we will further explore the degeneracy between metallicity
and dust depletion effects in these absorption systems by considering 
absorption-line measurements of interstellar zinc.

\subsection{Preliminary interpretation of observed trends}

The various trends presented in the previous sub-sections demonstrate that the 
Ca/Ti absorbers in our sample trace a broad range of gaseous environments in
the inner and outer regions of galaxies and in the circumgalactic environment. 
The sub-division on the absorbers into two different classes (class 1 and class 2) 
is justified in view of the different behavior of these two absorber classes 
concerning the observed Ca\,{\sc ii}/Ti\,{\sc ii} and Ti\,{\sc ii}/H\,{\sc i} ratios
(see previous section). Since Ti\,{\sc ii} and H\,{\sc i} have identical 
ionization potentials, the different trends for Ti\,{\sc ii}/H\,{\sc i} for the two
classes must originate in different dust-depletion properties and/or Ti abundances in the
two absorber classes. If dust is (primarily) responsible for the observed trends,
then class-1 absorbers trace optically thick gaseous regions in and around galaxies that 
have substantially less dust depletion of Ti than class-2 systems. The apparent lack of 
dust depletion in class-1 systems may be related to a different size or composition of dust 
grains in such environments (see, e.g., \citet{bdraine}) or simply caused by a lower 
dust-to-gas ratio. From the discussion in the previous section on the observed number density of 
Ti\,{\sc ii}-absorbing systems at low redshift follows that possibly half of the 
high-column density sub-DLAs and DLAs with log $N$(H\,{\sc i}$)>20.1$ may belong to
this category of absorbers.

Because of the steeper decline of [Ca/H] with H\,{\sc i} the effect of dust depletion in
the absorption systems is more severe for Ca, a trend that is well known for Milky Way 
gas (Savage \& Sembach 1996).
However, also photo-ionization affects the gas-phase abundance 
of singly-ionized Ca in interstellar gas, owing to the fact that the ionization 
potential of Ca\,{\sc ii} is smaller than that of H\,{\sc i}. 
For the interpretation of the observed Ca/Ti ratios in the
absorbers with respect to the nature and origin of the gas (for instance, disk gas vs. halo gas), 
a detailed photoionization modeling therefore is required. Such a modeling will be presented
in Sect.\,5.

%%%%%%%%%%%%%%%%%%%%%%%%%%%%%%%%%%%%%%%%%%%%%%%%%%%%%%%%%%%%%%%%%%%%%%%%%%%%%%%%%%%%%%%%%%%%%%%%%%%

\section{Comparison with Ca/Ti/H observations in other environments}

%%%%%%%%%%%%%%%%%%%%%%%%%%%%%%%%%%%%%%%%%%%%%%%%%%%%%%%%%%%%%%%%%%%%%%%%%%%%%%%%%%%%%%%%%%%%%%%%%%%

\begin{figure}
  \resizebox{\hsize}{!}{\includegraphics{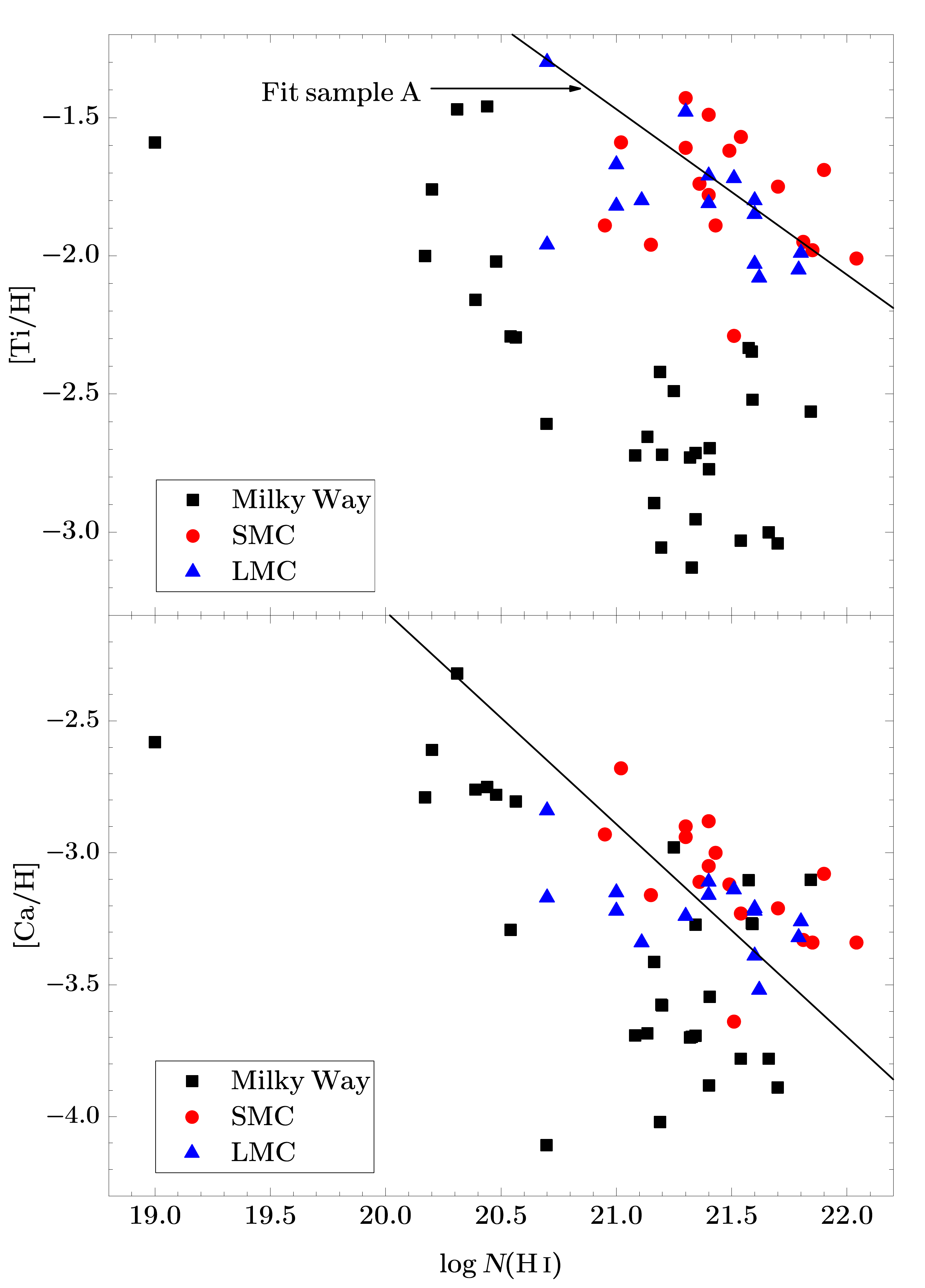}}
  \caption{$[\mathrm{Ti}/\mathrm{H}]$ and $[\mathrm{Ca}/\mathrm{H}]$ vs. $\log N(\mathrm{H\,\textsc{i}})$
  of absorption line systems in the Milky Way and the Small \& Large Magellanic Clouds (SMC and LMC,
  respectively).
  In case of $[\mathrm{Ti}/\mathrm{H}]$ the solid line represents the linear fit of the sample A systems,
  which in this case are class 1 systems, too. In case of $[\mathrm{Ca}/\mathrm{H}]$ the solid line
  represents the linear fit of all the QALS for which Ca\,\textsc{ii} as well as H\,\textsc{i} column
  densities are available.}
  \label{tihcahvshmw}
\end{figure}

%%%%%%%%%%%%%%%%%%%%%%%%%%%%%%%%%%%%%%%%%%%%%%%%%%%%%%%%%%%%%%%%%%%%%%%%%%%%%%%%%%%%%%%%%%%%%%%%%%%

\subsection{Milky Way and the Magellanic Clouds}

In the Milky Way ISM, the measured Ca\,{\sc ii}/Ti\,{\sc ii} ratios imply [Ti/Ca] ratios that are 
typically far below the ones seen in QALS, with a mean value of about $\langle$[Ti/Ca]$\rangle=-0.9$ 
(e.g., \citet{bhunter}).
The absolute depletion of Ti in the Milky Way ISM varies between $\sim 1.3$ dex for warm, diffuse
gas and $\sim 3.0$ for cold, dense gas, if the $\zeta$ Ophiuchi cloud is taken 
as reference (see Savage \& Sembach 1996).
In the Magellanic Clouds, the typical depletion of Ti into dust grains appears to be 
significantly smaller than in the Milky Way (see \citet{bwelty} and references therein), indicating a lower dust-to-gas ratio
and/or different grain properties.

When plotting $\log N$(Ti\,{\sc ii}) vs.\,$\log N$(Ca\,{\sc ii}) in our sample, class-1 absorbers follow the trend seen in the Magellanic Clouds, while the data points (=limits) for class-2 systems are in 
accordance with the trend seen in the Milky Way (see Fig.\,\ref{overview}).
To further compare the observed trends in QALS with those in the Milky Way (MW) and the Magellanic 
Clouds (MCs) in Fig.\,\ref{tihcahvshmw} we plot the abundance ratios [Ti/H] (upper panel) and 
[Ca/H] (lower panel) as a function of the neutral gas column density 
for the MW/MCs from \citet{bwelty} together with the
fitted trend seen for the intervening absorbers in sample A (Fig.\,4). 
Also here it is evident that the QALS in sample A are much closer to the SMC data points
than to the data points in the Milky Way,
indicating that the depletion of Ti and Ca into dust grains is substantially smaller 
in QSALs than in the Milky Way on a level that is comparable to the one in the SMC. 
This behavior most likely reflects a lower (average) dust-to-gas ratio
in these systems and/or a larger fraction (cross section) of warm, diffuse gas in which
the dust depletion is expected to be generally smaller than in cold, dense gas (see above). 

If most QALS typically have depletion properties similar to the SMC rather than similar 
to the Milky Way, this aspect needs to be considered in the interpretation of the 
observed abudances of dust-depleted elements in QALS.
In fact, as pointed out by \citet{bwelty}, comparing the abundance trends in intervening absorbers 
with the depletion pattern of heavy elements in the Milky Way ISM
may lead to inappropriate conclusions about the dust content and enrichment history of these
systems.

%%%%%%%%%%%%%%%%%%%%%%%%%%%%%%%%%%%%%%%%%%%%%%%%%%%%%%%%%%%%%%%%%%%%%%%%%%%%%%%%%%%%%%%%%%%%%%%%%%%

\begin{figure}
  \resizebox{\hsize}{!}{\includegraphics{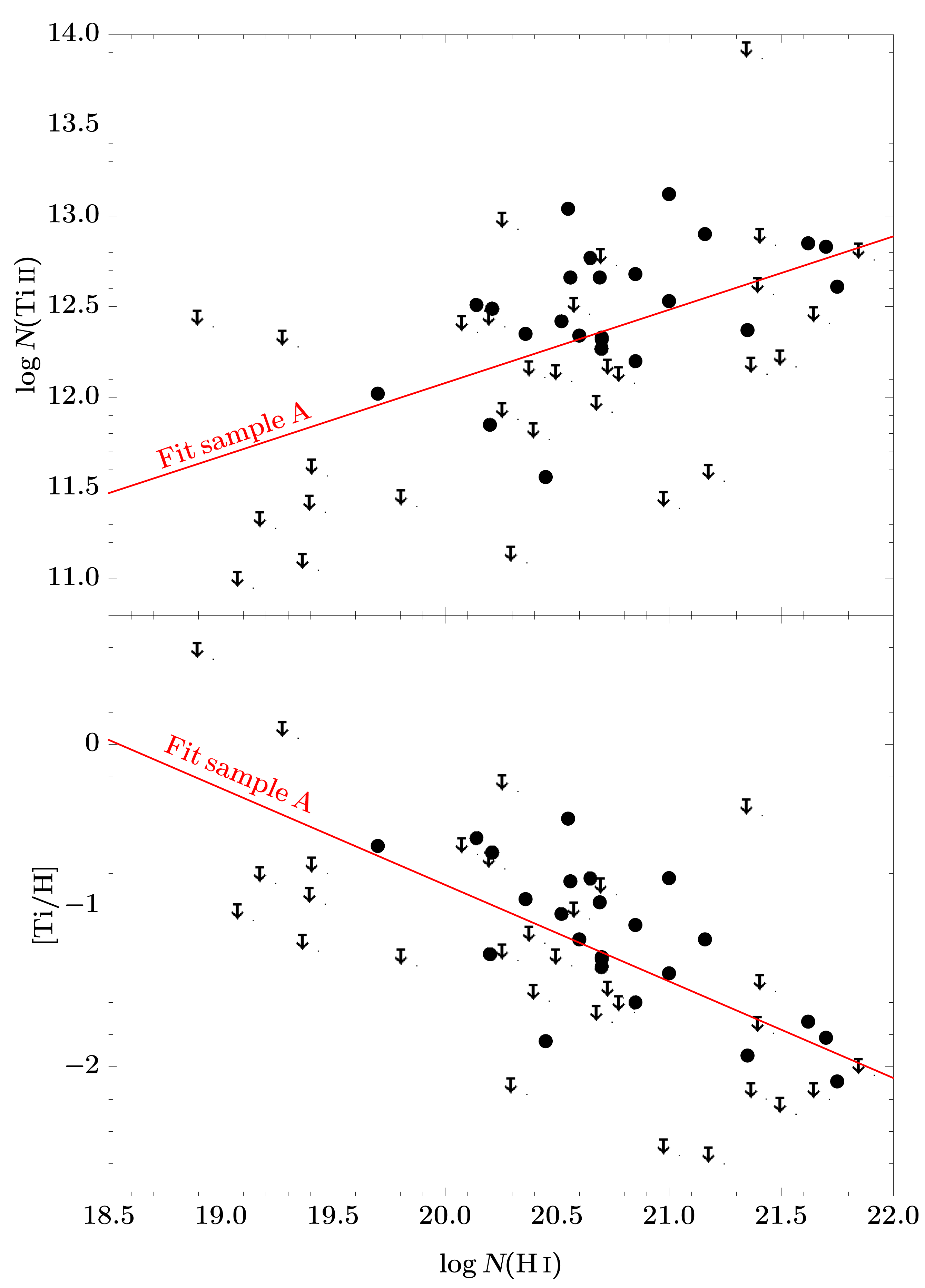}}
  \caption{$\log N(\mathrm{Ti\,\textsc{ii}})$ and $[\mathrm{Ti}/\mathrm{H}]$ vs. 
  $\log N(\mathrm{H\,\textsc{i}})$ for Ti\,\textsc{ii} systems without any information about 
  the Ca-content from Table\,\ref{tab clas 3}. The solid red lines represent the fits of the sample A/
  class 1 systems.}
  \label{Ticl3vsH}
  \end{figure}

%%%%%%%%%%%%%%%%%%%%%%%%%%%%%%%%%%%%%%%%%%%%%%%%%%%%%%%%%%%%%%%%%%%%%%%%%%%%%%%%%%%%%%%%%%%%%%%%%%%

\subsection{Other intervening absorption-line systems}

Systems for which no Ca column densities or upper limits are available obviously
cannot be used to study Ti/Ca ratios in intervening absorbers.
However, these systems can be used to test the above-stated hypothesis that
the Ti depletion is more severe for increasing metallicity of the gas.

\subsubsection{Ti-absorbers without Ca information}

In Fig.\,\ref{Ticl3vsH} we show the column densities for Ti\,{\sc ii} over H\,{\sc i} 
(upper panel) and [Ti/H] over H\,{\sc i} (lower panel) for the 51 predominantly high-redshift
absorbers from the list of \citet{bwelty}, for which there is no information on 
the Ca\,\textsc{ii} column density. These systems are listed in Table \,
\ref{tab clas 3} in the Appendix.
Although there is remarkable scatter in the plots, these high-redshift systems 
predominantly follow the same trends for Ti\,{\sc ii} over H\,{\sc i}
as the Ca\,{\sc ii}-selected systems from sample A/class 1, as presented in Sect.\,3.2 
(Fig.\,\ref{Ticl3vsH}, solid red lines). 
From this we conclude that the initial Ca\,{\sc ii} selection criterion for the low-redshift 
absorbers in our UVES sample apparently does not introduce a bias into the 
distribution of Ti\,{\sc ii}/H\,{\sc i} ratios in our absorber sample.

\subsubsection{General metallicity dependency of dust depletion}

To further investigate the relation between dust-depletion of Ca and Ti and
the overall metal abudance in the absorbers we consider supplementary
information on singly-ionized Zn (see Sect.\,2). The Zn\,{\sc ii} 
column densities are listed in Table\,\ref{tab clas 3} in the Appendix.

%%%%%%%%%%%%%%%%%%%%%%%%%%%%%%%%%%%%%%%%%%%%%%%%%%%%%%%%%%%%%%%%%%%%%%%%%%%%%%%%%%%%%%%%%%%%%%%%%%%

\begin{figure}
  \resizebox{\hsize}{!}{\includegraphics{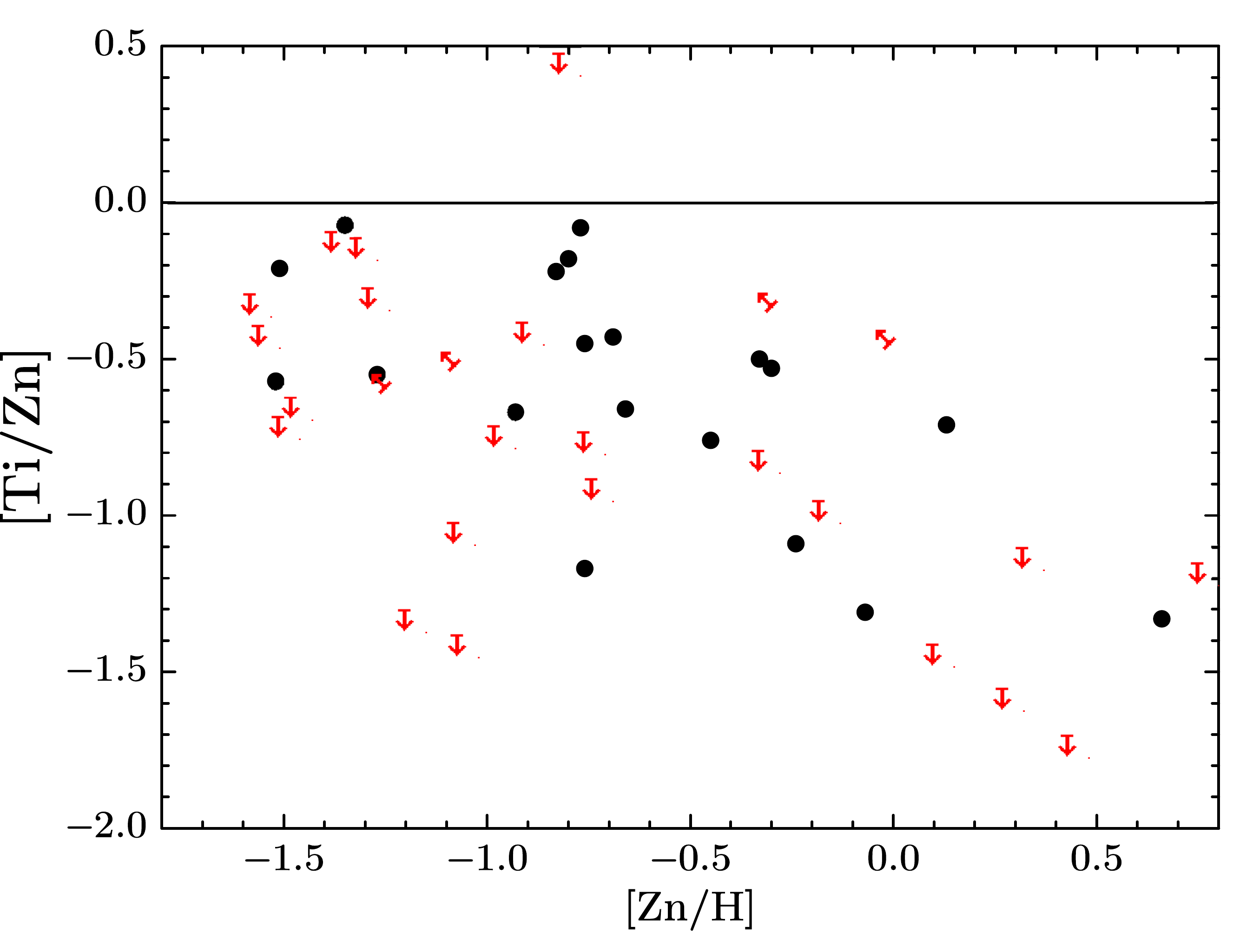}}
  \caption{$[\mathrm{Ti}/\mathrm{Zn}]$ vs. $[\mathrm{Zn}/\mathrm{H}]$. In contrast to Ti, the portion of Zn, which is depleted into dust grains can be expected to be comparatively small. Therefore $[\mathrm{Ti}/\mathrm{Zn}]$ can be treated as a good indicator of dust depletion. Also because of this $[\mathrm{Zn}/\mathrm{H}]$ is a good indicator for the metallicity of the observed system. This plot shows that on average dust depletion is more severe in systems with higher metallicity.}
  \label{TiZnvsZnH}
\end{figure}

\begin{figure}
  \resizebox{\hsize}{!}{\includegraphics{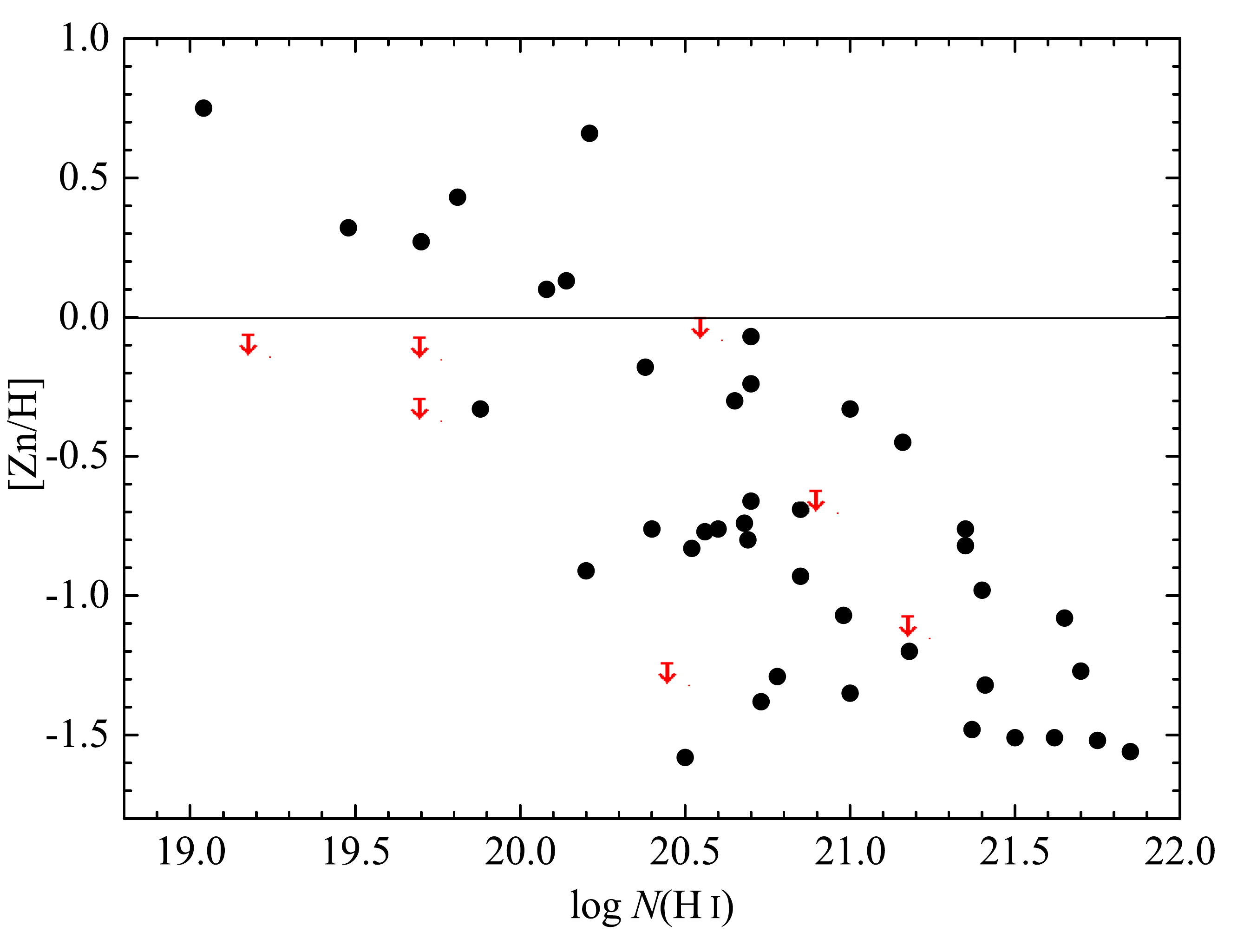}}
  \caption{Metallicity ($[\mathrm{Zn}/\mathrm{H}]$) as a function of neutral gas column density.}
  \label{znhvsh}
\end{figure}

%%%%%%%%%%%%%%%%%%%%%%%%%%%%%%%%%%%%%%%%%%%%%%%%%%%%%%%%%%%%%%%%%%%%%%%%%%%%%%%%%%%%%%%%%%%%%%%%%%%

\begin{figure}
  \resizebox{\hsize}{!}{\includegraphics{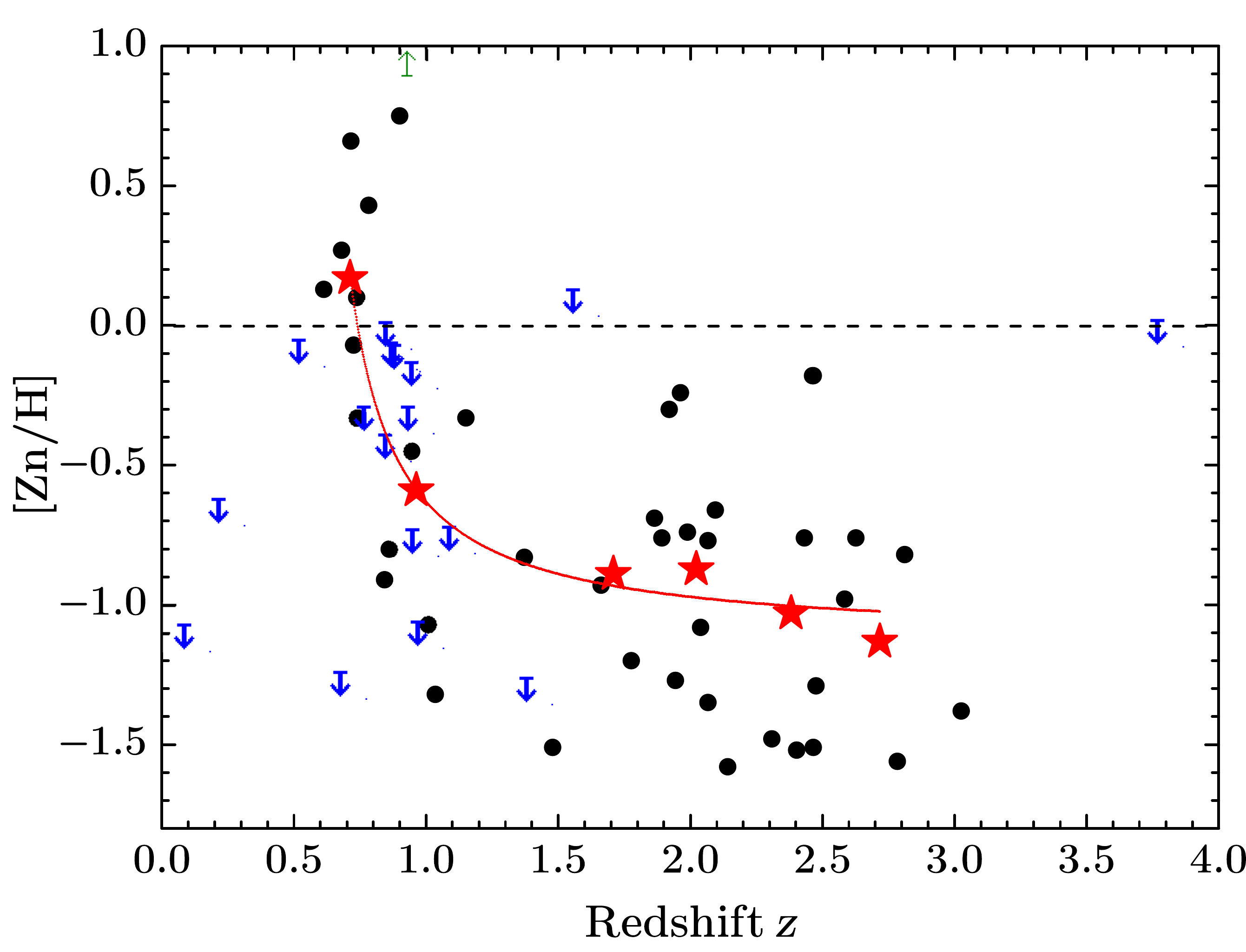}}
  \caption{$[\mathrm{Zn}/\mathrm{H}]$, as an indicator for the metallicity, vs. redshift $z$. Although as expected the scatter is remarkable, this plot indicates on average higher metallicities for younger systems.}
  \label{ZnHvsz}
\end{figure}

%%%%%%%%%%%%%%%%%%%%%%%%%%%%%%%%%%%%%%%%%%%%%%%%%%%%%%%%%%%%%%%%%%%%%%%%%%%%%%%%%%%%%%%%%%%%%%%%%%%

\begin{figure}
  \resizebox{\hsize}{!}{\includegraphics{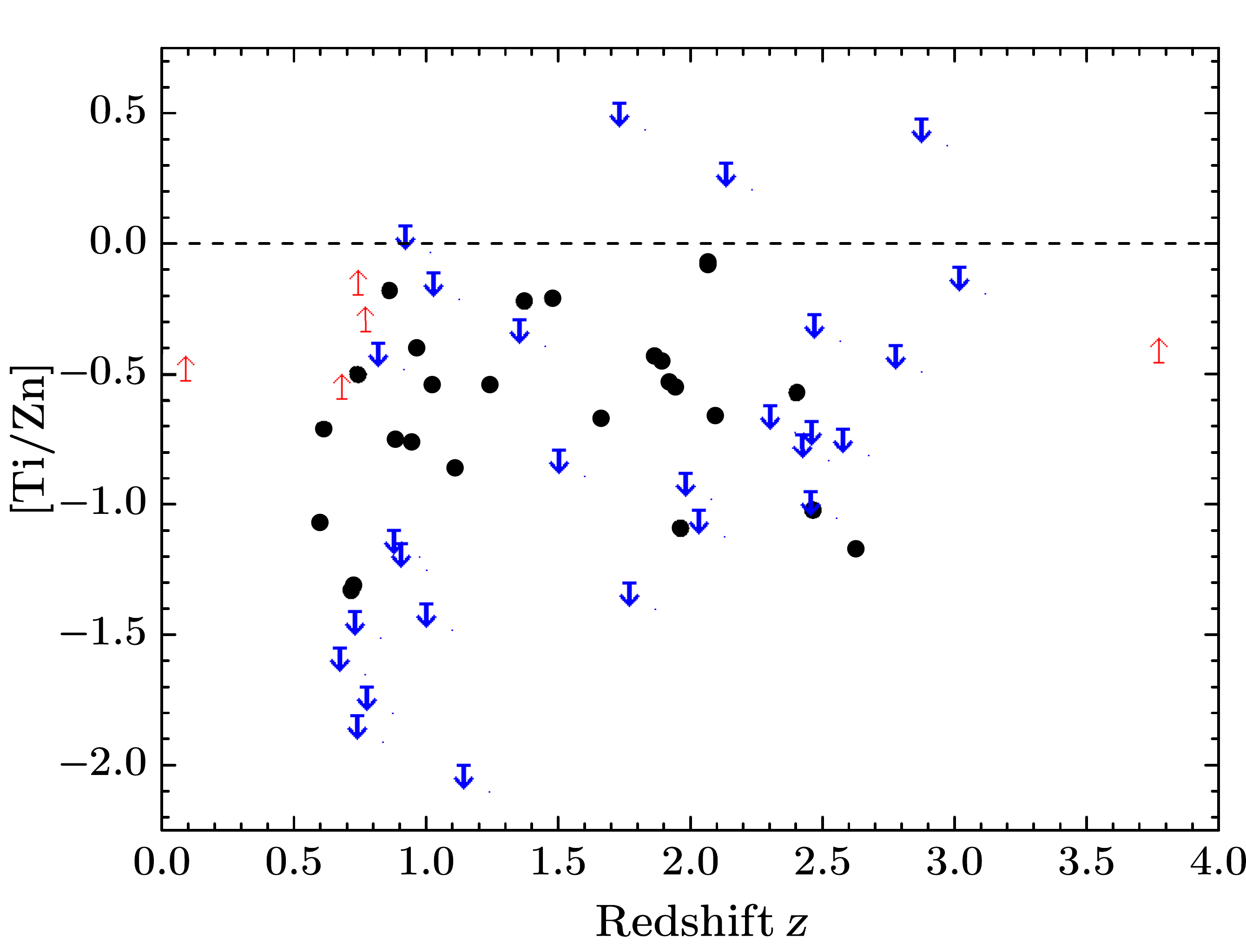}}
  \caption{$[\mathrm{Ti}/\mathrm{Zn}]$ vs. redshift $z$. In contrast to $[\mathrm{Zn}/\mathrm{H}]$ (Fig.\, \ref{ZnHvsz}), there`s no comparable trend visible for $[\mathrm{Ti}/\mathrm{Zn}]$ as a function of $z$. }
  \label{TiZnvsz}
\end{figure}

%%%%%%%%%%%%%%%%%%%%%%%%%%%%%%%%%%%%%%%%%%%%%%%%%%%%%%%%%%%%%%%%%%%%%%%%%%%%%%%%%%%%%%%%%%%%%%%%%%%

In Fig.\,\ref{TiZnvsZnH} we plot the $[\mathrm{Ti}/\mathrm{Zn}]$ ratio as a function of the 
Zn abundance $[\mathrm{Zn}/\mathrm{H}]$ (i.e., the metallicity) of the absorbers. 
The trend seen in this figure indicates that the depletion of Ti increases for increasing 
metallicity, possibly because of an increase of the dust-to-gas ratio for increasing metallicity.
The scatter is substantial but not surprising, because dust depletion not only depends on the 
overall metallicity of the gas, but also on other physical parameters, such as 
densities, temperatures and the strengths of the interstellar radiation fields (e.g., \citet{bdraine}).

Fig.\,\ref{znhvsh} shows a clear trend for decreasing metallicity with increasing 
H\,\textsc{i} column density. For DLAs this trend is well-known (e.g., \citet{bkhare}).
Both dependencies put together lead to a weak trend of less severe dust depletion 
for increasing hydrogen column density. 

For Galactic sightlines exactly the opposite trend is observed \citep{bwelty}.
This is because the absorbing gas clouds in the Milky Way have similar metallicities 
and so indeed depletion is more severe for increasing $N(\mathrm{H}\,\textsc{i})$, 
because, on average, gas volume densities increase with increasing $N(\mathrm{H}\,\textsc{i})$
\footnote{In a 5 dimensional phase space with the axes dust depletion $D(\mathrm{X})$, $n(\mathrm{H})$, 
$T$, strength of radiation field and metallicity all systems we expect to lay on a 
4 dimensional hyper surface.}.
In our sample of extragalactic absorbers, in contrast, we sample multiple arbitrary 
sightlines through  {\it different} (unrelated) optically thick gas environments and 
each absorber (i.e., each data point) is characterized by its own metallicity and dust 
properties (sightline averaged). The distribution of data points for the intervening 
data points further depends strongly on the cross section of the absorbing gas environments.
Obviously, low-metallicity, high-column density neutral gas regions have a larger 
absorption cross section than high-metallicy environments with 
somewhat lower neutral gas columns.

Unfortunately, the derived parameters for the class 1 systems include column densities for either 
H\,\textsc{i} {\it or} Zn\,\textsc{ii}, but not both species together. 
Only for one class 1 system we know $N(\mathrm{H\,\textsc{i}})$ and a lower limit for 
$N(\mathrm{Zn\,\textsc{ii}})$. 
%Neglecting the one outlier with $[\mathrm{Ti}/\mathrm{Ca}]=-0.08$, all other high Ti/Ca systems show comparable values for $[\mathrm{Ca}/\mathrm{H}]$, $[\mathrm{Ca}/\mathrm{Zn}]$, $[\mathrm{Ti}/\mathrm{H}]$ and $[\mathrm{Ti}/\mathrm{Zn}]$ each.
%[PLEASE CHECK THE FOLLOWING TEXT AND MODIFY/UPDATE CALCULATIONS ACCORDINGLY.]
%In particular, 
Table\, \ref{tab class 1 and 2} indicates that class 1 systems 
are characterized by $[\mathrm{Ti}/\mathrm{Zn}]$ ratios that are larger than $-0.8$,
while class 2 systems have $[\mathrm{Ti}/\mathrm{Zn}]\leq-0.8$, typically. 
If we consider the $[\mathrm{Ti}/\mathrm{Zn}]$ ratio as discriminator
also for the 57 absorbers without Ca\,{\sc ii} information in Table\,\ref{tab clas 3},
we separate 18 systems with $[\mathrm{Ti}/\mathrm{Zn}]>-0.8$ (sample C) from those 12 systems
with $[\mathrm{Ti}/\mathrm{Zn}]\leq-0.8$ (sample D). The 27 systems for which 
there are only limits on $[\mathrm{Ti}/\mathrm{Zn}]$ that do not allows us to
distinguish between sample C and D are collected in sample E. 
%(UPDATE LAST COLUMN OF THAT TABLE AND SPLIT IN SAMPE C,D,E).

The mean Zn abundance in sample C absorbers is $\left<[\mathrm{Zn}/\mathrm{H}]\right>=-0.80 \pm 0.47$,
which is mildly %(substantially?) 
lower than the mean Zn abundance in sample D 
$\left<[\mathrm{Zn}/\mathrm{H}]\right>=-0.37\pm 0.65$, although the uncertainties are substantial.
\textbf{The only one value for the Zn abundance of a class 1 system is comparatively low ($[\mathrm{Zn}/\mathrm{H}]<-1.08$), while for Class 2 systems it is comparatively high ($\langle [\mathrm{Zn}/\mathrm{H}]\rangle= 0.04$).} 
Thus, if absorbers in sample A (=class 1) and sample C have similar dust 
properties (as suggested by the similarly high $[\mathrm{Ti}/\mathrm{Zn}]$ ratios)
and both samples belong to class 1 systems, the observed trend is in line with our previous
suggestion, namely that class 1 systems typically represent metal- and dust-poor gas 
environments where the Ti depletion is much less severe than for Ca, so that 
the observed Ti/Ca ratios are relatively high.
Although our data do not confirm metallicity as the only parameter determining the dust depletion, the observed systematic differences in $[\mathrm{Ti}/\mathrm{Zn}]$ demonstrate that both samples (C \& D) general exhibit explicitly different dust properties.

\subsubsection{Redshift dependency of abundance and depletion properties}

Here we shortly discuss the redshift dependency of the metallicity and 
titanium dust depletion in our absorber sample. In Fig.\,\ref{ZnHvsz} we 
have plotted the Zn abundance in the individual absorbers as a function of 
$z$ (black filled circles). The distribution of data points exhibits 
substantial scatter.
To explore the global trend we binned every seven systems, calculated the 
average values of the metallicity $([\mathrm{Zn}/\mathrm{H}])$ and redshift and
fitted a reciprocal function (red stars and red solid line).
One can see that, on average, the metallicity decreases with increasing redshift
(see also \citet{brafelski}).

While the mean Zn abundance in the absorbers decreases with 
increasing redshift, Fig.\,\ref{TiZnvsz} shows that Ti dust depletion
(as indicated by the Ti/Zn ratio) has no striking redshift dependence,
but instead scatters with $\sim 2.5$ dex. There is a weak tendency for 
very low-redshift systems having (on average) a lower Ti/Zn ratio (i.e.,
a larger Ti depletion), which would be in line with the higher average
metallicity of low-$z$ and the presumably higher dust content. This 
trend becomes significant, if the upper limits for Ti/Zn (blue arrows) are 
taken into account.

Such a large scatter for Ti/Zn is not surprising because the dust content is not
only a function of the global metallicity of the absorbers (which itself exhibits
a huge spread; see above), but also depends on other parameters such as local gas density, 
temperature, and strength of the interstellar radiation field.

%%%%%%%%%%%%%%%%%%%%%%%%%%%%%%%%%%%%%%%%%%%%%%%%%%%%%%%%%%%%%%%%%%%%%%%%%%%%%%%%%%%%%%%%%%%%%%%%%%%

\begin{figure}
  \resizebox{\hsize}{!}{\includegraphics{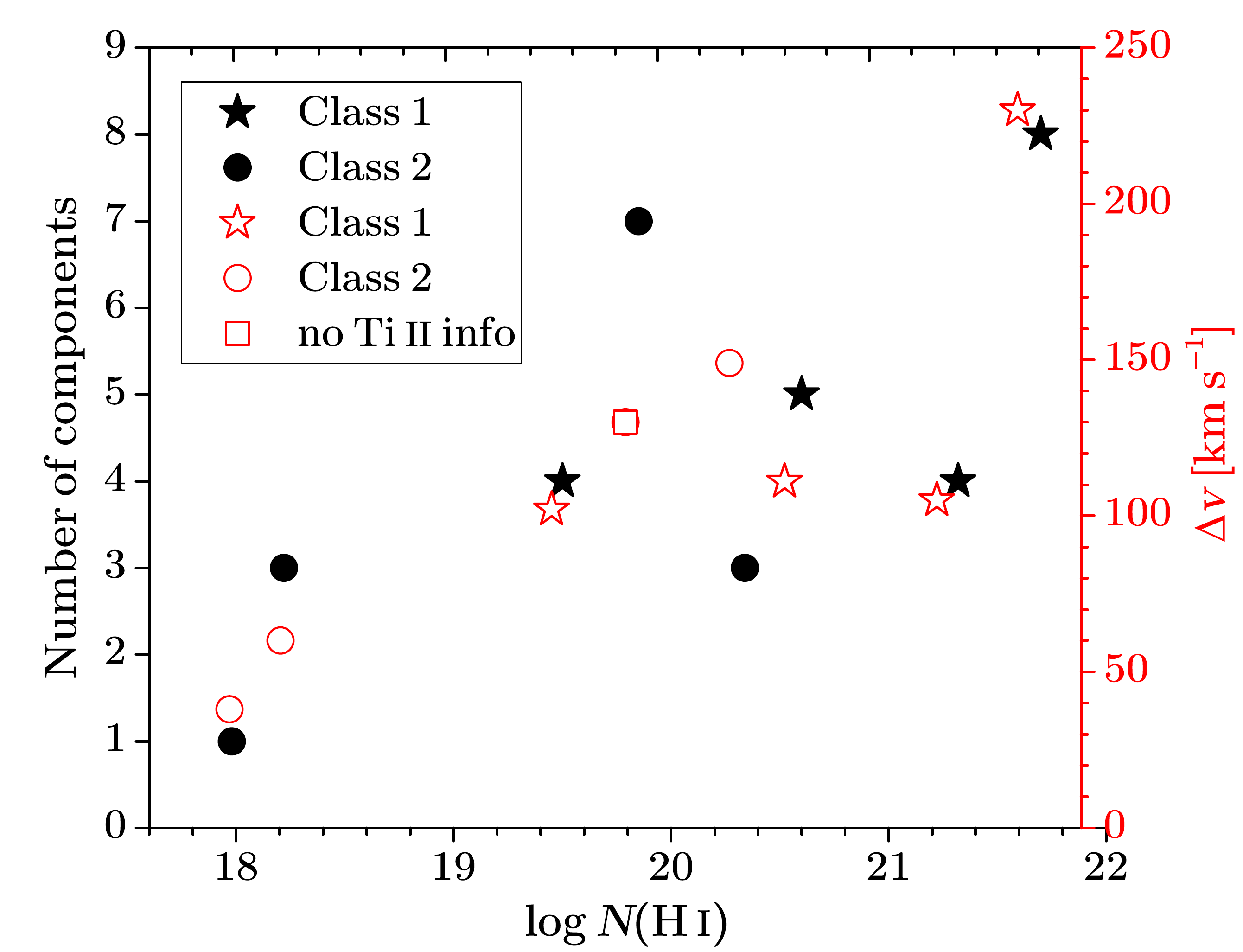}}
  \caption{Number of components and rest frame velocity width $\Delta v$ 
  of the Ca\,\textsc{ii} $\lambda3934$ absorption feature vs. 
  $\log N(\mathrm{H\,\textsc{i}})$ for class 1 \& 2 systems.} 
  \label{nrdvvsh}
\end{figure}

%%%%%%%%%%%%%%%%%%%%%%%%%%%%%%%%%%%%%%%%%%%%%%%%%%%%%%%%%%%%%%%%%%%%%%%%%%%%%%%%%%%%%%%%%%%%%%%%%%%

\subsubsection{The role of absorber kinematics and impact parameters}

Fig.\,\ref{nrdvvsh} shows that for most of the systems for which information on the number 
of components ($N_{\rm comp.}$), the rest-frame velocity width of the Ca\,\textsc{ii} absorption ($\Delta v$),
as well as the H\,\textsc{i} column density is available there is a correlation between these parameters: 
$\Delta v$ increases with an increasing component number and both parameters increase for increasing 
$\log N(\mathrm{H\,\textsc{i}})$. These trends hold for both, class 1 and class 2 systems.
Obviously, the (predominantly neutral) gas phase traced by Ca\,\textsc{ii} is kinematically 
more complex in high-column density systems than in low-column density absorbers 
(see also \citet{brichter}).

Less clear is the dependence of the parameters $N_{\rm comp.}$, $\Delta v$, 
$\log N(\mathrm{H\,\textsc{i}})$, [Ca/H], and [Ti/Ca] on the impact parameters $d$ of the 
absorbers to their host galaxies (Fig.\,\ref{allinall}). 
Only for a few systems information on $d$ is available, so that we refrain from drawing 
any conclusions from these plots.

\subsubsection{Notes on nucleosynthetic aspects}

Out of the three examined elements, Ca, Ti, and Zn, only Ca is a true alpha
element, meaning that the most abundant isotope $^{40}\mathrm{Ca}$, with a natural
relative abundance of $\sim 97$ \%, is a direct result of the so called alpha-process.
Alpha elements or isotopes, respectively, are built up by the addition of
$\upalpha$-particles during the last phase of a massive stars life, and their
nuclei consist of an integer multiple of an $\upalpha$-particle. Alpha
elements mainly are produced during or short before core-collapse supernovae.
Although not a direct member of the alpha process, many authors
(e.g., \citet{bwelty}) claim an abundance pattern for Ti similar to $\alpha$ elements.
On the other hand, zinc and iron trace each other (\citet{btimmes} and references therein),
making Zn an element with an abundance pattern similar as an iron peak element.
These elements typically are formed during thermonuclear fusions in Type-1 supernovae.

As a consequence, not only the overall (absolute) metallicities of the examined 
absorption systems typically are substantially different from the local ISM value, but 
also the {\it relative} abundances of Ca, Ti, and Zn may be other than what is indicated
by the solar reference values \citep{basplund}. 
This aspect needs to be kept in mind when interpreting the column
densities of the above listed ions in intervening absorption-line systems.

\section{Ionization modeling}

\subsection{Model setup} 

As mentioned earlier, the ionization potential of Ca\,\textsc{ii} is smaller than that 
of H\,\textsc{i}. As a result, in neutral regions the amount of Ca\,\textsc{ii} in the 
gas phase might be much smaller than the total amount of Ca.
To investigate ionization effects for singly-ionized Ca (and for the other two
ions, Ti\,{\sc ii} and Zn\,{\sc ii}) and their impact on the observed 
ion ratios in our absorbers we modeled the ionization conditions in the gas using 
ionization code \texttt{Cloudy C13.1} \citep{bferland}. 
Gas absorbers are modeled as plane-parallel slabs of constant neutral columns 
that are illuminated by an external ionizing radiation field.
The big unknown in the modeling setup is the shape and intensity of the UV radiation
field that critically determines the ionization balance between Ca\,\textsc{ii} and  Ca\,\textsc{iii} 
in diffuse neutral gas. \texttt{Cloudy} offers several options to pre-define the 
ionizing radiation field. 
We here consider two different modeling setups with different radiation fields,
which we refer to ``halo-model'' and ``disk-model'' in the following.
For both models we select the following initial conditions for our \texttt{Cloudy} grids:

\begin{itemize}
\item Cosmic microwave background at $z=0.5$,
\item Cosmic ray background,
\item Gas temperature grid with $T/\SI{e+3}{\kelvin} \in \lbrace 0.1, 0.5, 1, 2.5, 5, 10, 20\rbrace$,
\item H\,{\sc i} column-density grid with log [$N$(H\,{\sc i})/cm$^{-2}]
      \in \lbrace19.5, 20.5, 21.5, 22.5\rbrace$,
\item Solar reference abundances given by \citet{basplund}, and
\item Standard ISM-grain properties (\texttt{grains ISM} command).
\end{itemize}

Note that included \texttt{grains ISM} option does {\it not} affect the gas-phase 
abundances of the elements included in the dust grains, i.e., dust {\it depletion}
effects are not included in the model. \textbf{Also note that the true relative abundances may differ from the solar values and also the dust grain composition may be different from the composition assumed via the \texttt{grains ISM} command. This may lead to further uncertainties in the assumed model.}

The ionizing radiation field in the disk model is based on the local, unextinguished interstellar 
Galactic radiation field of \citet{bblack} (via \texttt{table ISM} command). 
The halo-model, in contrast, assumes the Haardt \& Madau (2005) extragalactic 
UV background field (via the \texttt{table HM05} command) normalized at $z=0.3$. 
For comparison, the spectral energy distribution of both background fields 
are shown in Fig.\,\ref{radfields} in the Appendix.

%%%%%%%%%%%%%%%%%%%%%%%%%%%%%%%%%%%%%%%%%%%%%%%%%%%%%%%%%%%%%%%%%%%%%%%%%%%%%%%%%%%%%%%%%%%%%%%%%%%

\begin{figure}
  \resizebox{\hsize}{!}{\includegraphics{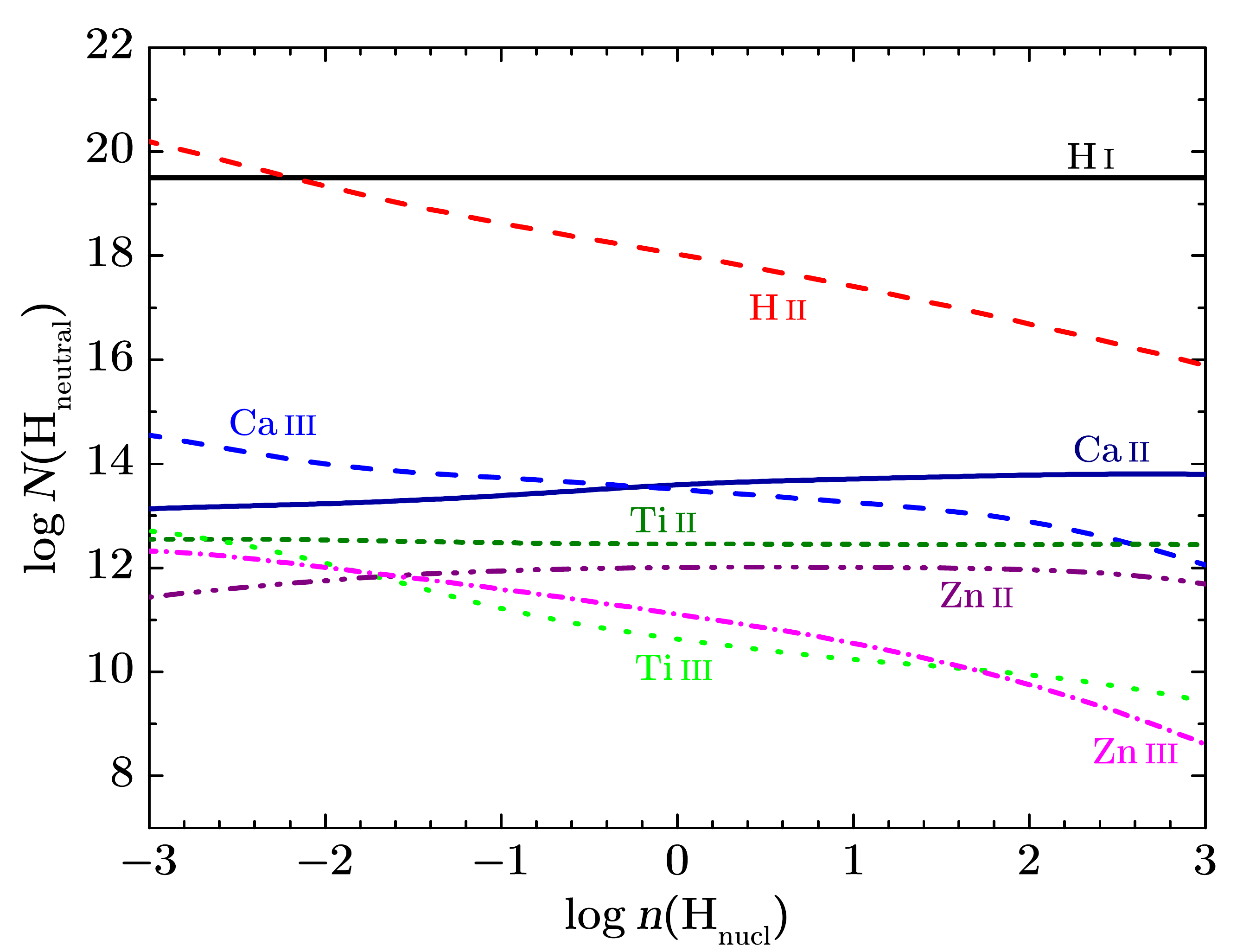}}
  \caption{Example of a \texttt{Cloudy}-model output: logarithm of (ion)-column densities as a 
  function of the log of the hydrogen density for a halo-like system with 
  log $N$(H\,{\sc i}$)=19.5$ and a temperature of 
  \SI{5000}{\kelvin}. The column densities $N$ are given in \SI{}{\per\square\centi\metre}, 
  the density $n$ in \SI{}{\per\cubic\centi\metre}.}
  \label{grid3example}
\end{figure}

%%%%%%%%%%%%%%%%%%%%%%%%%%%%%%%%%%%%%%%%%%%%%%%%%%%%%%%%%%%%%%%%%%%%%%%%%%%%%%%%%%%%%%%%%%%%%%%%%%%

\subsection{Modeling results}

As an illustration for a typical \texttt{Cloudy} output in Fig.\,\ref{grid3example} we show the model
predictions for the column densities of the relevant ionization states of H, Ca, Ti, and Zn 
as a function of the gas density for an absorber that is exposed to the extragalactic UV
field (halo-model).

%%%%%%%%%%%%%%%%%%%%%%%%%%%%%%%%%%%%%%%%%%%%%%%%%%%%%%%%%%%%%%%%%%%%%%%%%%%%%%%%%%%%%%%%%%%%%%%%%%%

\begin{figure}
  \resizebox{\hsize}{!}{\includegraphics{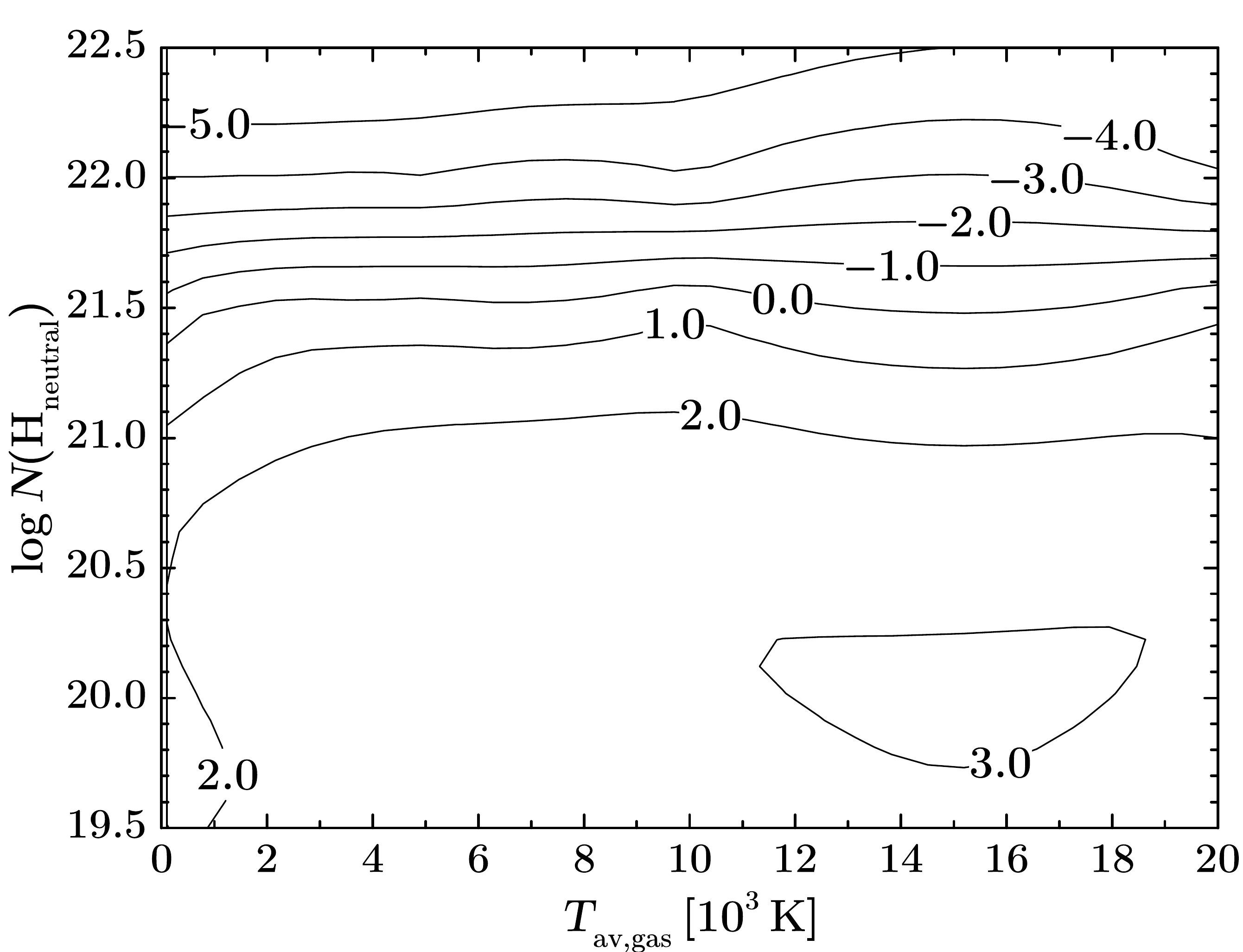}}
  \caption{\texttt{Cloudy} result for the disk model grid: contour plot of the critical hydrogen density 
  log $n_\mathrm{crit}$ above which $N(\mathrm{Ca}\,\textsc{ii}) \geq N(\mathrm{Ca}\,\textsc{iii})$
  in the $N(H_\mathrm{neutral})/T$ space.}
  \label{grid2result}
\end{figure}

%%%%%%%%%%%%%%%%%%%%%%%%%%%%%%%%%%%%%%%%%%%%%%%%%%%%%%%%%%%%%%%%%%%%%%%%%%%%%%%%%%%%%%%%%%%%%%%%%%%

\begin{figure}
  \resizebox{\hsize}{!}{\includegraphics{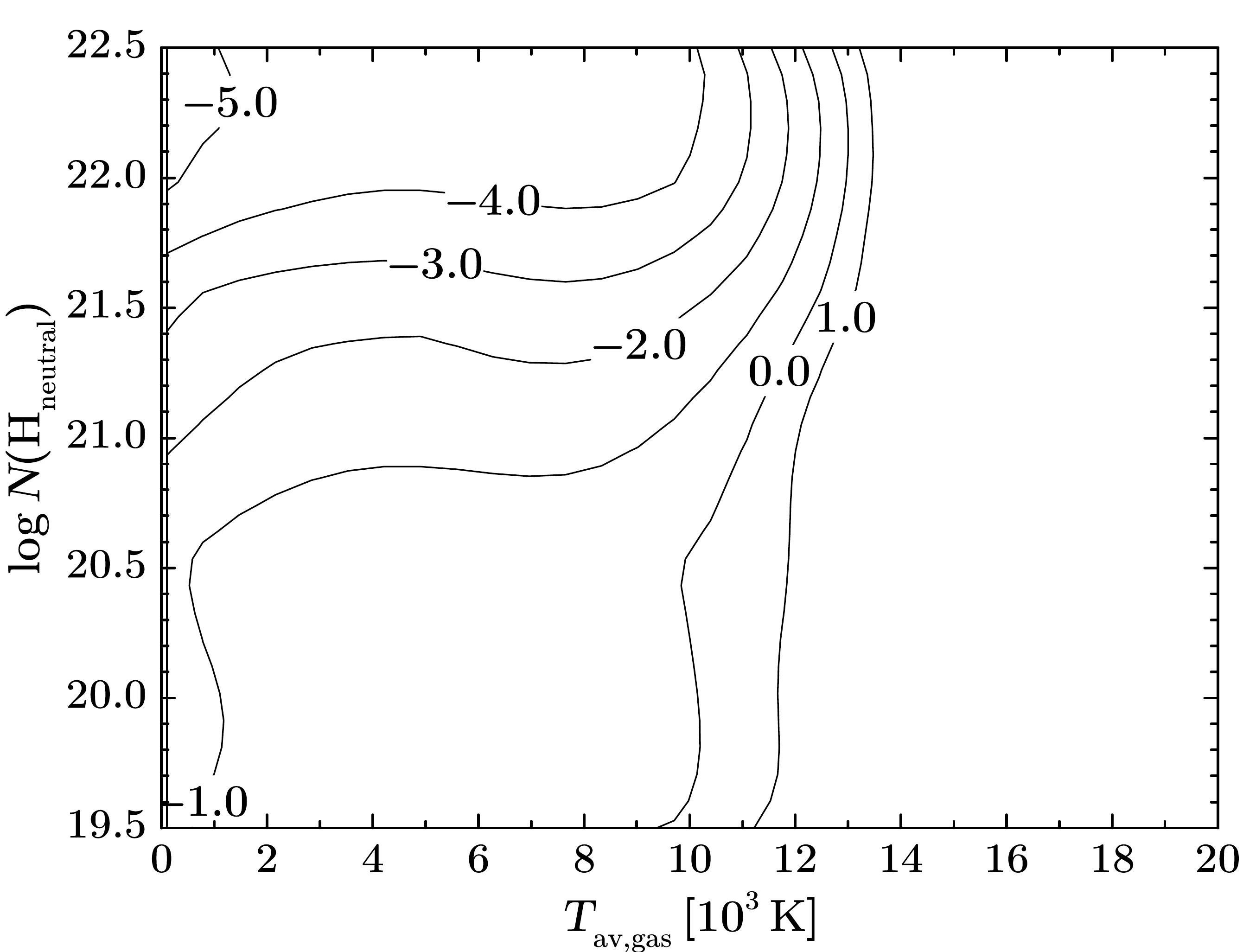}}
  \caption{\texttt{Cloudy} result for the halo model grid: 
  contour plot of the critical hydrogen density
  log $n_\mathrm{crit}$ above which $N(\mathrm{Ca}\,\textsc{ii}) \geq N(\mathrm{Ca}\,\textsc{iii})$
  in the $N$(H\,{\sc i})/$T$ space.
  Above \SI{12000}{\kelvin} \texttt{Cloudy} only delivers solutions for which 
  $N(\mathrm{Ca}\,\textsc{ii})<N(\mathrm{Ca}\,\textsc{iii})$ for all 
  $\log n(\mathrm{H}_\mathrm{nucl})$.}
  \label{grid3result}
\end{figure}

%%%%%%%%%%%%%%%%%%%%%%%%%%%%%%%%%%%%%%%%%%%%%%%%%%%%%%%%%%%%%%%%%%%%%%%%%%%%%%%%%%%%%%%%%%%%%%%%%%%

For the two model grids (disk model, halo model) we investigate, under which conditions 
the considered metal ions (singly-ionized Ca, singly-ionized Ti, singly-ionized Zn) represent
the dominant ionization states in the gas. In particular, we constrain the 
critical hydrogen density log $n_\mathrm{crit}$ that separates from each other
the dominant ionization states for H, Ca, Ti, and Zn in their 
characteristic gas phases.
In Figs.\,\ref{grid2result} and \ref{grid3result} we show 
the resulting values for $n_\mathrm{crit}$ for Ca\,{\sc ii} as contours 
in the $N$(H\,{\sc i})/$T$ phase space.

The main results from the \texttt{Cloudy} calculation for the two model-grids can be
summarized as follows:\\
\\
\noindent
{\it Disk model}:
the critical hydrogen density $n_\mathrm{crit}$ for 
which $N(\mathrm{H}\,\textsc{i})=N(\mathrm{H}\,\textsc{ii})$ and 
$N(\mathrm{Ca}\,\textsc{ii})=N(\mathrm{Ca}\,\textsc{iii})$, respectively, 
turns out to be independent of the gas temperature. 
For hydrogen, $\log n_\mathrm{crit}$ increases smoothly with decreasing 
$N$(H\,{\sc i}). For Ca, in contrast,  $\log n_\mathrm{crit}$ 
falls steeply for log $N$(H\,{\sc i}$)>21.5$. The disk model indicates
that hydrogen is predominantly neutral for all log $n_{\rm H}>0$. 
Ca\,\textsc{ii} is expected to be the dominant Ca ion for all 
log $N$(H\,{\sc i}$)>22$, while Ca\,\textsc{iii} is dominant 
for all log $N$(H\,{\sc i}$)<21.5$.
Ti\,\textsc{ii} or Zn\,\textsc{ii} represent the dominant Ti- and 
Zn-ions, respectively, in the entire parameter phase space. 
\\
\noindent
{\it Halo model}:
for gas temperatures above \SI{1.5e+4}{\kelvin} hydrogen is predominantly ionized, 
while for temperatures below \SI{1.2e+4}{\kelvin} hydrogen is predominately neutral.
Only for temperatures below \SI{12e+3}{\kelvin} and densities log $n_{\rm H}>0$ 
Ca\,\textsc{ii} is expected to be the dominant ionization state of Ca in 
these environments, while Ca\,\textsc{iii} dominates for gas that is warmer 
and/or more diffuse. As for the disk model, Ti\,\textsc{ii} or Zn\,\textsc{ii} 
are the dominant ions in the entire parameter phase space.

In conclusion, the \texttt{Cloudy} model grids suggest that the interpretation
of Ca\,\textsc{ii} column densities and ion ratios that include Ca\,\textsc{ii}
in individual absorbers is problematic without knowing the exact local physical 
conditions. For singly-ionized Ti and Zn, in contrast, the local physical 
conditions are less important and thus the observed Ti\,\textsc{ii} and Zn\,\textsc{ii} 
column densities provide robust measures for the gas-phase abundances of these 
elements in predominantly neutral gas regions.

Despite these theoretical uncertainties for the interpretation of Ca\,\textsc{ii} 
(either as dominant or non-dominant ion) in individual intervening absorption systems, 
the {\it observed} relation between 
Ca\,\textsc{ii}/H\,{\sc i} and H\,\textsc{i} (Fig.\,\ref{TiCavsH}) indicates that
the analysis of Ca\,\textsc{ii} in a large-enough sample does provide useful information
on the dust depletion properties in such systems if compared to other observable 
ions that co-exist with singly-ionized Ca in the same gas phase.

%%%%%%%%%%%%%%%%%%%%%%%%%%%%%%%%%%%%%%%%%%%%%%%%%%%%%%%%%%%%%%%%%%%%%%%%%%%%%%%%%%%%%%%%%%%%%%%%%%%

\section{Summary}

In this paper we have studied dust depletion of Ti and Ca in gas in and around galaxies 
by analyzing Ti/Ca abundance ratios in intervening absorption-line systems at 
low and high redshift.

First, we have investigated high-resolution optical spectra obtained by UVES/VLT 
and spectroscopically analyzed 34 absorption-line systems
at $z_\mathrm{abs}\leq 0.5$ to measure column densities (or limits) for Ca\,{\sc ii} and Ti\,{\sc ii}.
Secondly, We have complemented our UVES data set with previously published 
absorption-line data on Ti/Ca for redshifts up to $z\approx 3.8$ and created an 
absorber sample containing more than 100 absorbers including DLAs, sub-DLAs, and LLS.

Our study suggests that there are two distinct populations of Ti/Ca absorbing systems with
either high or low Ti/Ca ratios. We refer to these two populations as ``class 1'' and ``class 2''
absorbers, respectively,

Class 1 absorbers are characterized by 
i) relatively high Ti/Ca-ratios $([\mathrm{Ti}/\mathrm{Ca}]\gtrsim+1.0)$,
ii) relatively low metallicities $([\mathrm{Zn}/\mathrm{H}]\lesssim-1.0)$,
iii) and relatively mild Ti-dust-depletion values $([\mathrm{Ti}/\mathrm{Zn}]>-0.8)$.
In class 1 absorbers, $\mathrm{Ti}/\mathrm{Ca}]$ appears to increase with 
increasing $N(\mathrm{H\,\textsc{i}})$, implying that dust depletion becomes
progressively more important for Ca compared to Ti in systems with large 
neutral gas columns. The simultaneous significant detection of both ions, 
Ca\,\textsc{ii} \& Ti\,\textsc{ii}, is a good indicator for observing a class 1 
system. Class 1 systems obviously represent metal-poor gas absorbers
with relatively low dust content.

In contrast to class 1 systems, class 2 absorbers typically have higher metallicities
$([\mathrm{Zn}/\mathrm{H}]\gtrsim 0.0)$ and a higher dust-to-gas ratio. Here, the 
Ti depletion reaches similar values as for Ca, leading to systemaically lower 
Ti/Ca-ratios $([\mathrm{Ti}/\mathrm{Ca}]\lesssim+1.0)$.
The class 2 systems do not show any systematic relation between H\,{\sc i}
column density and the ion ratios $[\mathrm{Ti}/\mathrm{H}]$ and 
$[\mathrm{Ti}/\mathrm{Ca}]$. 

From our \texttt{Cloudy} ionization model grids we conclude 
that Ti\,\textsc{ii} \& Zn\,\textsc{ii} are the dominant ionization states 
of Ti \& Zn in neutral gas within a broad range of physical conditions.
For Ca, however, the \texttt{Cloudy} models predict a strong dependency
of the ionization fraction of singly- and double-ionized Ca on the 
local physical conditions. Therefore, ionization effects and dust depletion
effects are difficult to disentangle for Ca\,\textsc{ii} in individual 
absorbers without information on other depleted and undepleted species.

The Ti/Ca ratios in class 1 systems are very similar to those observed in the
LMC and SMC, but are substantially higher than what is found in the Milky Way ISM. 
This trend is in line with the above stated conclusion that class 1 systems with high Ti/Ca
ratios trace dust- and metal-deficient gas environments in and around galaxies.

Our study suggests that the Ti/Ca ratio in intervening absorption systems may be 
used as indicator for the metal- and dust-content of intervening absorbers. 
This might be of relevance in particular for those low-redshift systems for 
which direct information on the metallicity cannot be obtained due to the 
lack of UV spectral data that would give access to the Lyman series of H\,{\sc i}.
In addition, future studies on Ti/Ca in intervening absorbers {\it in combination} with 
supplementary data on galaxy impact parameters for a large-enough absorber/galaxy 
sample hold the prospect to boost our understanding of the distribution 
of dust in the inner and outer regions of galaxies.

%% CHECK in class 1 systems Ti/Ca high becuase of IONZATION effects? Mostly  dust, as Ca follows
%%% same trand as other depleted elements for which ionization effects are less improtant.

%%%%%%%%%%%%%%%%%%%%%%%%%%%%%%%%%%%%%%%%%%%%%%%%%%%%%%%%%%%%%%%%%%%%%%%%%%%%%%%%%%%%%%%%%%%%%%%%%%%

\appendix

\section{Additional Figures and Tables}
\newpage
\begin{sidewaystable*}
\centering
\scriptsize
\caption{Data for Ca\,{\sc ii} systems with known Ti\,{\sc ii} column density (SAMPLE A) or upper 
limits for Ti\,{\sc ii} (SAMPLE B). A star symbol in front of the sightline marks systems analyzed in this 
paper. Data of all other systems are from table A3 in \citet{bwelty}. 
With one exception all sample A systems are also class 1 systems $([\mathrm{Ti}/\mathrm{Ca}]\gtrsim+1)$. 
The sample A system with $z=1.0232$ is a class 2 system $[\mathrm{Ti}/\mathrm{Ca}]\simeq0)$. For 
most of the sample B systems the upper limit of $[\mathrm{Ti}/\mathrm{Ca}]$ is too high to decide 
whether it is a class 1 or 2 system. We expect systems with $[\mathrm{Ti}/\mathrm{Ca}]<0.5$ to be 
class 2 systems.}
\label{tab class 1 and 2}
\begin{tabular}{lllrrrrrrrrrr}
\hline\hline
Class
&QSO / sightline
&$z_\mathrm{abs}$
&$\log N($H\,{\sc i}$)$
&$\log N($Ca\,{\sc ii}$)$
&$\log N($Ti\,{\sc ii}$)$
&$\log N($Zn\,{\sc ii}$)$
&$[\mathrm{Ca}/\mathrm{H}]$
&$[\mathrm{Ca}/\mathrm{Zn}]$
&$[\mathrm{Ti}/\mathrm{H}]$
&$[\mathrm{Ti}/\mathrm{Zn}]$
&$[\mathrm{Ti}/\mathrm{Ca}]$
&$[\mathrm{Zn}/\mathrm{H}]$\\
\hline
\multicolumn{11}{c}{SAMPLE A: Ca\,{\sc ii} systems with known Ti\,{\sc ii} column density}\\
\hline
1       &Q0738$+$313    &0.0912 &21.18  &12.32  &12.53  &$<12.66$       &$-3.2$ &$>-2.12$       &$-1.6$ &$>-0.52$       &1.6    &$<-1.08$       \\
1       &*J095456$+$174331      &0.23782        &$21.32 \pm 0.07^a$     &$12.211 \pm 0.023$     &$12.565 \pm 0.046$     &$-$    &$-3.45 \pm 0.14$       &$-$    &$-1.71 \pm 0.17$       &$-$    &$1.74 \pm 0.16$\\
1       &*J235731$-$112539      &0.24763        &$-$    &$12.849 \pm 0.033$     &$12.917 \pm 0.067$     &$-$    &$-$    &$-$    &$-$    &$-$    &$1.46 \pm 0.19$        &$-$\\
1       &*J113007$-$144927      &0.31273        &$21.71 \pm 0.09^b$     &$12.7142 \pm 0.0062$   &$12.782 \pm 0.021$     &$-$    &$-3.33 \pm 0.14$       &$-$    &$-1.88 \pm 0.16$       &$-$    &$1.46 \pm 0.12$        &$-$\\
1       &*J123200$-$022404      &0.39498        &$20.6^c$ &$12.398 \pm 0.014$     &$12.363 \pm 0.040$     &$-$    &$-2.542$       &$-$    &$-1.187$       &$-$    &$1.36 \pm 0.15$        &$-$\\
1       &*J045608$-$215909      &0.47439        &$19.5^d$ &$12.23 \pm 0.014$      &$11.825 \pm 0.046$     &$-$    &$-1.61$        &$-$    &$-0.625$       &$-$    &$0.99 \pm 0.15$\\
1       &Q0118$-$272    &0.558  &$20.3^e$       &12.37  &12.27  &$-$    &$-2.27$        &$-$    &$-0.98$        &$-$    &1.29   &$-$\\
1       &J0846$+$0529   &0.7429 &$-$    &13.06  &13.00  &$<12.8$        &$-$    &$>-1.52$       &$-$    &$>-0.19$       &$1.33$ &$-$\\
1       &J1007$+$2853   &0.8839 &$-$    &13.33  &13.13  &13.49  &$-$    &$-1.94$        &$-$    &$-0.75$        &1.19   &$-$\\
1       &J1129$+$0204   &0.9650 &$-$    &13.11  &12.79  &12.8   &$-$    &$-1.47$        &$-$    &$-0.4$ &1.07   &$-$\\
2       &J0953$+$0801   &1.0232 &$-$    &13.57  &12.10  &12.25  &$-$    &$-0.46$        &$-$    &$-0.54$        &$-0.08$        &$-$\\
1       &J1430$+$0149   &1.2418 &$-$    &12.85  &12.79  &12.94  &$-$    &$-1.87$        &$-$    &$-0.54$        &1.33   &$-$    \\
\hline
\multicolumn{11}{c}{SAMPLE B: Ca\,{\sc ii} systems with upper limits for the Ti\,{\sc ii} column density}\\
\hline
2       &*J121509$+$330955      &0.00396        &$20.34^f$        &$12.366 \pm 0.018$     &$<11.361$      &$-$    &$-2.314$       &$-$    &$<-1.93$      &$-$    &$<0.39$        &$-$\\
2       &*J133007$-$205616      &0.01831        &$-$    &$13.049 \pm 0.025$     &$<11.817$      &$-$    &$-$    &$-$    &$-$    &$-$    &$<0.16$ &$-$\\
?       &*J215501$+$092224      &0.08091        &$17.98^g$        &$11.861 \pm 0.047$     &$<11.182$      &$-$    &$-0.46$        &$-$    &$<0.26$       &$-$    &$<0.72$        &$-$\\
?       &Q0738$+$313    &0.2210 &20.90  &11.91  &$<11.480$      &$<12.83$       &$-3.33$        &$>-2.7$        &$<-2.37$       &$-$    &$<0.96$        &$<-0.63$\\
?       &*J012517$-$001828      &0.23864        &$-$    &$11.447 \pm 0.092$     &$<11.298$      &$-$    &$-$    &$-$    &$-$    &$-$    &$<1.24$        &$-$\\
2       &*J000344$-$232355      &$0.27051$      &$-$    &$11.660 \pm 0.025$     &$<10.827$      &$-$    &$-$    &$-$    &$-$    &$-$    &$<0.56$        &$-$\\
?       &*J142249$-$272756      &0.27563        &$-$    &$12.135 \pm 0.012$     &$<11.392$      &$-$    &$-$    &$-$    &$-$    &$-$    &$<0.65$        &$-$\\
?       &*J042707$-$130253      &0.28929        &$-$    &$11.626 \pm 0.061$     &$<11.378$      &$-$    &$-$    &$-$    &$-$    &$-$    &$<1.15$        &$-$\\
?       &*J102837$-$010027      &0.32427        &$-$    &$12.496 \pm 0.029$     &$<11.679$      &$-$    &$-$    &$-$    &$-$    &$-$    &$<0.58$        &$-$\\
?		&*J231359$-$370446		&0.33980		&$-$	&$12.542 \pm 0.030$		&$<12.190$		&$-$	&$-$	&$-$	&$-$	&$-$	&$<1.04$\\
?       &*J142253$-$000149      &0.34468        &$-$    &$12.535 \pm 0.046$     &$<11.872$      &$-$    &$-$    &$-$    &$-$    &$-$    &$<0.73$        &$-$\\
?       &*J110325$-$264515      &0.35896        &$-$    &$11.251 \pm 0.039$     &$<10.541$      &$-$    &$-$    &$-$    &$-$    &$-$    &$<0.68$        &$-$\\
?       &*J121140$+$103002      &0.39293        &$-$    &$11.388 \pm 0.108$     &$<11.208$      &$-$    &$-$    &$-$    &$-$    &$-$    &$<1.21$        &$-$\\
?       &*J050112$-$015914      &0.40310        &$-$    &$12.350 \pm 0.048$     &$<12.075$      &$-$    &$-$    &$-$    &$-$    &$-$    &$<1.12$        &$-$\\
2       &*J224752$-$123719      &0.40968        &$-$    &$12.268 \pm 0.037$     &$<11.337$      &$-$    &$-$    &$-$    &$-$    &$-$    &$<0.46$        &$-$\\
?       &*J232820$+$002238      &0.41277        &$-$    &$11.521 \pm 0.073$     &$<11.138$      &$-$    &$-$    &$-$    &$-$    &$-$    &$<1.01$        &$-$\\
?       &*J051707$-$441055      &0.42913        &$-$    &$10.476 \pm 0.059$     &$<10.192$      &$-$    &$-$    &$-$    &$-$    &$-$    &$<1.11$        &$-$\\
?       &*J220743$-$534633      &0.43720        &$-$    &$12.002 \pm 0.069$     &$<11.498$      &$-$    &$-$    &$-$    &$-$    &$-$    &$<0.89$        &$-$\\
?       &*J044117$-$431343      &0.44075        &$18.22 \pm 0.20^h$     &$11.482 \pm 0.061$     &$<11.197$      &$-$    &$-1.01 \pm 0.30$       &$-$    &$<0.026$       &$-$    &$<1.11$        &$-$\\
?       &*J144653$+$011356      &0.44402        &$-$    &$11.338 \pm 0.110$     &$<10.864$      &$-$    &$-$    &$-$    &$-$    &$-$    &$<0.92$        &$-$\\
?       &*J124646$-$254749      &0.49282        &$-$    &$12.773 \pm 0.079$     &$<11.922$      &$-$    &$-$    &$-$    &$-$    &$-$    &$<0.54$        &$-$\\
2       &Q2335$+$1501   &0.6798 &19.70  &12.41  &$<11.36$       &12.53  &$-1.63$        &$-1.9$ &$<-1.29$       &$<-1.56$       &$<0.34$        &0.27\\
2       &Q1436$-$0051   &0.7377 &20.08  &12.83  &$<11.71$       &12.74  &$-1.59$        &$-1.69$        &$<-1.32$       &$<-1.42$       &$<0.27$        &0.1\\
2       &J1203$+$1028   &0.7463 &$-$    &13.05  &$<11.20$       &$12.63$        &$-$    &$-1.36$        &$-$    &$<-1.82$       &$<-0.46$       &$-$\\
?       &Q1009$-$0026   &0.8866 &19.48  &12.26  &$<11.64$       &12.36  &$-1.56$        &$-1.88$        &$<-0.79$       &$<-1.11$       &$<0.77$        &0.32\\
?       &Q1631$+$1156   &0.9004 &19.70  &12.17  &$<11.66$       &$<12.18$       &$-1.87$        &$>-1.79$       &$<-0.99$       &$-$    &$<0.88$        &$<-0.08$\\
?       &Q0826$-$2230   &0.9110 &19.04  &11.75  &$<11.58$       &12.35  &$-1.63$        &$-2.38$        &$<-0.41$       &$<-1.16$       &$<1.22$        &0.75\\
?       &Q1436$-$0051   &0.9281 &$<18.80$       &12.28  &$<12.71$       &12.26  &$>-0.86$       &$-1.76$        &$-$    &$<0.06$        &$<1.82$        &$>0.90$\\
?       &J0517$-$441    &1.1496 &$-$    &12.74  &$<10.6$        &$12.22$        &$-$    &$-1.26$        &$-$    &$<-2.01$       &$<-0.75$       &$-$\\
2       &HE0515$-$4414  &1.151  &19.88  &12.72  &$<11.7$        &12.11  &$-1.5$ &$-1.17$        &$<-1.13$       &$<-0.80$       &$<0.37$        &$-0.33$\\
\hline
\end{tabular}
%\begin{tablenotes}

\medskip
\footnotesize
\textbf{References}. (a) \citet{brao}; (b) \citet{blane}; (c) \citet{blebrun}; (d) \citet{bturnshek}; 
(e) \citet{bvladilo}; (f) \citet{bmiller}; (g) \citet{bjenkins}; (h) Value calculated using the observed 
equivalent width of the H$\alpha$ line given by \citet{bryabinkov} $(\SI{1.53\pm0.11}{\angstrom})$ with 
the assumption that this line is in the damping part of the curve of growth.
%\end{tablenotes}
\end{sidewaystable*}

\begin{table*}
 \centering
  \caption{Data for Ti\,{\sc ii} systems or upper limit Ti\,{\sc ii} systems without any Ca\,{\sc ii} information. 
 Redshifts and column densities are taken from Welty \& Crowther (2010; see references therein). For calculation of ratios solar abundances given by \citet{basplund} are assumed.}
  \label{tab clas 3}
  \begin{tabular}{llrrrrrr}
  \hline\hline
QSO &$z_\mathrm{abs}$&$\log N($H\,{\sc i}$)$	&$\log N($Ti\,{\sc ii}$)$		&$\log N($Zn\,{\sc ii}$)$ &$[\mathrm{Ti}/\mathrm{H}]$ &$[\mathrm{Ti}/\mathrm{Zn}]$ &$[\mathrm{Zn}/\mathrm{H}]$\\
\hline
\multicolumn{8}{c}{SAMPLE C: $[\mathrm{Ti}/\mathrm{Ca}]>-0.8$}\\
\hline
Q0058$+$019		&0.612	&20.14		&12.51		&12.83		&$-0.58$	&$-0.71$	&0.13		\\
Q1122$-$168		&0.682	&20.45		&11.56		&$<11.76$	&$-1.84$	&$>-0.59$	&$<-1.25$\\
J1107$+$0048	&0.7403	&21.00		&13.12		&13.23		&$-0.83$	&$-0.50$	&$-0.33$	\\
Q0153$+$0009	&0.7714	&19.70		&12.02		&$<11.96$	&$-0.63$	&$>-0.33$	&$<-0.3$\\
Q0454$+$039		&0.8598	&20.69		&12.66 		&12.45		&$-0.98$	&$-0.18$	&$-0.80$\\
J1727$+$5302	&0.9449	&21.16		&12.90		&13.27		&$-1.21$	&$-0.76$	&$-0.45	$\\
Q0935$+$417		&1.3726	&20.52			&12.42		&12.25		&$-1.05$	&$-0.22$	&$-0.83$\\
Q0933$+$733		&1.4790	&21.62			&12.85		&12.67		&$-1.72$	&$-0.21$	&$-1.51$\\
Q1104$-$181		&1.662	&20.85			&12.20		&12.48		&$-1.60$	&$-0.67$	&$-0.93$\\
Q2230$+$025		&1.8644	&20.85			&12.68		&12.72		&$-1.12$	&$-0.43$	&$-0.69$\\
Q1210$+$17		&1.8918	&20.60			&12.34		&12.40		&$-1.21$	&$-0.45$	&$-0.76$\\
Q2206$-$199		&1.920	&20.65			&12.77		&12.91		&$-0.83$	&$-0.53$	&$-0.30$\\
Q1157$+$014	&1.944	&21.70			&12.83		&12.99		&$-1.82$	&$-0.55$	&$-1.27$\\
Q2231$-$002	&2.0661	&20.56			&12.66		&12.35		&$-0.85$	&$-0.08$	&$-0.77$\\
F0812$+$32	&2.0668	&21.00			&12.53		&12.21		&$-1.42$	&$-0.07$	&$-1.35$\\
Q2359$-$02	&2.0951	&20.70			&12.33		&12.60		&$-1.32$	&$-0.66$	&$-0.66$\\
Q0027$-$1836	&2.402	&21.75		&12.61		&12.79		&$-2.09$	&$-0.57$	&$-1.52$\\
P0133$+$0400	&3.7736	&20.55		&13.04		&$<13.1$	&$-0.46$	&$>-0.45$	&$<-0.01$\\
\hline
\multicolumn{8}{c}{SAMPLE D: $[\mathrm{Ti}/\mathrm{Ca}]\leq 0.8$}\\
\hline
J0334$-$0711	&0.5976	&$-$		&11.9		&12.58		&$-$	&$-1.07$	&$-$	\\
J1323$-$0021	&0.716	&20.21		&12.49		&13.43		&$-0.67$	&$-1.33$	&0.66\\
J0256$+$0110	&0.725	&20.70		&12.27		&13.19		&$-1.38$	&$-1.31$	&$-0.07$\\
Q0138$-$0005	&0.7821	&19.81		&$<11.48$	&12.8		&$<-1.28$	&$<-1.71$	&0.43\\
Q0449$-$1645	&1.0072	&20.98		&$<11.47$	&12.47		&$<-2.46$	&$<-1.39$	&$-1.07$\\
Q0014$+$813		&1.11	&$-$		&12.36		&12.83		&$-$	&$-0.86$ &$-$\\
Q1331$+$170		&1.7765	&21.18			&$<11.62$	&12.54		&$<-2.51$	&$<-1.31$	&$-1.2$\\
Q0551$-$364	&1.962	&20.70			&12.32		&13.02		&$-1.33$	&$-1.09$	&$-0.24$\\
Q2318$-$1107	&1.989	&20.68		&$<12.00$	&12.50		&$<-1.63$	&$<-0.89$	&$-0.74$\\
Q0458$-$020	&2.0395	&21.65			&$<12.49$	&13.13		&$<-2.11$	&$<-1.03$	&$-1.08$\\
Q0201$+$365	&2.4628	&20.38			&$<12.19$	&12.76		&$<-1.14$	&$<-0.96$	&$-0.18$\\
Q0836$+$113	&2.4651	&20.58			&$<12.54$	&$<12.12$	&$<-0.99$	&$-1.02$	&$-0.18$\\
\hline
\multicolumn{8}{c}{SAMPLE E: no information about $[\mathrm{Ti}/\mathrm{Ca}]$ or upper limit too high}\\
\hline
Q2128$-$123		&0.430	&19.37 		&$<11.13$	&$-$		&$<-1.19$	&$-$&$-$	\\
Q0827$+$243		&0.5247	&20.30 		&$<11.76$	&$<12.8$	&$<-1.49$	&$-$ &$<-0.06$	\\
Q1622$+$238		&0.656	&20.36		&12.35		&$-$		&$-0.96$	&$-$	&$-$	\\
Q2206$-$199		&0.752	&$-$		&12.20		&$-$		&$-$		&$-$	&$-$	\\
Q1009$-$0026	&0.8426	&20.20		&11.85		&$<11.85$	&$<-1.30$	&$<-0.39$	&$-0.91$\\
Q0005$+$0524	&0.8514	&19.08		&$<11.03$	&$<11.24$	&$<-1.00$	&$-$	&$<-0.4$\\
Q1330$-$2056	&0.8526	&19.40		&$<11.45$	&$<11.96$	&$<-0.90$	&$-$	&$<0.00$\\
Q1228$+$1018	&0.9376	&19.41		&$<11.65$	&$<11.67$	&$<-0.71$	&$-$	&$<-0.30$\\
Q1054$-$0020	&0.9514	&19.28		&$<12.36$	&$<11.70$	&$<0.13$	&$-$	&$<-0.14$\\
Q1107$+$0003	&0.9547	&20.26		&$<13.01$	&$<12.08$	&$<-0.20$	&$-$	&$<-0.74$\\
Q1220$-$0040	&0.9746	&20.20		&$<12.47$	&$<11.69$	&$<-0.68$	&$-$	&$<-1.07$\\
Q1455$-$0045	&1.0929	&20.08		&$<12.44$	&$<11.91$	&$<-0.59$	&$-$	&$<-0.73$\\
Q0453$-$423		&1.150	&$-$		&$<13.11$	&$<12.63$	&$-$	&$-$	&$-$\\
J2240$-$0053	&1.3606	&$-$		&$<12.71$	&12.62		&$-$	&$<-0.30$	&$-$\\
Q0012$-$0122	&1.3862	&20.26		&$<11.96$	&$<11.55$	&$<-1.25$	&$-$	&$<-1.27$\\
Q0427$-$1302	&1.5623	&18.90		&$<12.47$	&$<11.58$	&$<0.62$	&$-$	&$<0.12$\\
Q1946$+$766		&1.7382	&$-$			&$<12.45$	&11.53		&$-$	&$<0.53$	&$-$	\\
Q0149$+$33	&2.1408	&20.50			&$<12.17$	&11.48		&$<-1.28$	&$<0.30$	&$-1.58$\\
Q0100$+$130	&2.309	&21.37			&$<12.21$	&12.45		&$<-2.11$	&$<-0.63$	&$-1.48$\\
Q2343$+$123	&2.4313	&20.40			&$<11.85$	&12.20		&$<-1.5$	&$<-0.74$	&$-0.76$\\
Q1223$+$175	&2.4661	&21.50			&$<12.25$	&12.55		&$<-2.20$	&$<-0.69$	&$-1.51$\\
Q0841$+$129	&2.4762	&20.78			&$<12.16$	&12.05		&$<-1.57$	&$<-0.28$	&$-1.29$\\
Q1209$+$0919	&2.5841	&21.40		&$<12.65$	&12.98		&$<-1.70$	&$<-0.72$	&$-0.98$\\
P1253$-$0228	&2.7828	&21.85		&$<12.84$	&12.85		&$<-1.96$	&$<-0.40$	&$-1.56$\\
Q0528$-$251	&2.8110	&21.35			&$<13.95$	&13.09		&$<-0.35$	&$<0.47$	&$-0.82$\\
Q0347$-$382	&3.0247	&20.73			&$<12.20$	&11.91		&$<-1.48$	&$<-0.10$	&$-1.38$\\
J0255$+$00	&3.2529 &20.70			&$<12.81$	&$-$		&$<-0.84$	&$-$	&$-$\\
\hline
\end{tabular}
\end{table*}

\begin{table*}
%\begin{sidewaystable*}
\centering
\small
\caption{Data for systems with only upper limits for both, Ca\,{\sc ii} AND Ti\,{\sc ii}, and data for systems with  information only for one ion or systems not matching one of the conditions mentioned before. For no one of the listed systems the values $[\mathrm{Ca}/\mathrm{Zn}]$, $[\mathrm{Ti}/\mathrm{Zn}]$ and $[\mathrm{Ti}/\mathrm{Ca}]$ are calculable. A star symbol in front of the sightline marks systems analyzed in this paper.}
\label{tab data cl 4}
\begin{tabular}{llrrrrrrr}
\hline\hline
QSO 
&$z_\mathrm{abs}$
&$\log N($H\,{\sc i}$)$	
&$\log N($Ca\,{\sc ii}$)$	
&$\log N($Ti\,{\sc ii}$)$		
&$\log N($Zn\,{\sc ii}$)$	
&$[\mathrm{Ca}/\mathrm{H}]$
&$[\mathrm{Ti}/\mathrm{H}]$
&$[\mathrm{Zn}/\mathrm{H}]$\\
\hline
\multicolumn{9}{c}{Systems with only upper limits both, for Ca\,{\sc ii} AND Ti\,{\sc ii}}\\
\hline
Q0827$+$243	&0.2590	&$-$	&$<11.60$	&$<11.96$	&$-$	&$-$	&$-$	&$-$\\
*J162439$+$234512	&0.31759	&$-$	&$<11.551$	&$<11.481$	&$-$	&$-$	&$-$	&$-$\\
*J005102$-$252846	&0.34393	&$-$	&$<11.293$	&$<12.057$	&$-$	&$-$	&$-$	&$-$\\
*J055158$-$211949	&0.43982	&$-$	&$<11.875$	&$<11.584$	&$-$	&$-$	&$-$	&$-$\\
*J030844$-$295702	&0.441	&$-$	&$<11.417$	&$<11.365$	&$-$	&$-$	&$-$	&$-$\\
*J000344$-$232355	&0.45241	&$-$	&$<10.466$	&$<10.635$	&$-$	&$-$	&$-$	&$-$\\
*J014125$-$092843	&0.50053	&$-$	&$<10.961$	&$<11.309$	&$-$	&$-$	&$-$	&$-$\\
Q2352$-$0028	&0.8730	&19.18	&$<11.01$	&$<11.36$	&$<11.67$	&$<-2.51$	&$<-0.77$	&$<-0.07$\\
\hline
\multicolumn{9}{c}{Systems without any information about Ti\,{\sc ii}  and other systems}\\
\hline
*J044117$-$431343	&0.10114	&$19.85 \pm 0.10^a$	&$12.606 \pm 0.013$	&ood	&$-$	&$-1.57 \pm 0.16$	&$-$	&$-$\\
*J104733$+$052454	&0.31822	&$-$				&$<10.853$			&ood	&$-$	&$-$					&$-$	&$-$\\
*J094253$-$110426	&0.39098	&$-$	&$11.929 \pm 0.019$	&blended	&$-$	&$-$	&$-$	&$-$\\
\hline
\end{tabular}

 \medskip
 \textbf{References.} (a) \citet{bchen}
%\end{sidewaystable*}
\end{table*}

\begin{table*}
\centering
 \caption{Data relating the absorption structure complexity and the absorber galaxy connection. Redshifts, component number and velocity widths of the Ca\,{\sc ii} $\lambda3934$ are from \citet{brichter}. The last column includes the references for impact parameters (calculated from angles using $\lambda$CDM cosmology and the cosmological parameters from \citet{bplanck} listed in Section 2.2.}
 \label{tab abs gal conn}
 \begin{tabular}{llrrrr}
  \hline\hline
	QSO sightline	& $z_{\mathrm{abs}}$	&Comp. nr. 	&$\Delta v_{\mathrm{CaII}\,\lambda3934}$ &Imp. par. $d$	&Ref.\\
					&		&			&$[\SI{}{\kilo\meter\per \second}]$	&$[\mathrm{kpc}]$	&	\\
  \hline
	\multicolumn{5}{c}{Sample A / Class 1}\\
  \hline
  J095456+174331	&0.23782	&4	&105	&$<4.8$	&a\\
  J235731$-$112539	&0.24763	&3	&53		&$-$	&\\
  J113007$-$144927	&0.31273	&8	&230	&18.5	&b\\
  J123200$-$022404	&0.39498	&5	&111	&6.6	&c\\
  J045608$-$215909	&0.47439	&4	&102	&$-$	&\\
  Q0118-272			&0.558		&$-$	&$-$	&10.7	&d\\
  \hline
  \multicolumn{5}{c}{Sample B}\\
  \hline
  J121509+330955	&0.00396	&3	&149	&9.7	&e\\
  J133007$-$205616	&0.01831	&9	&378	&$-$	&\\
  J215501+092224	&0.08091	&1	&38		&36.5	&f\\
  J012517$-$001828	&0.23864	&1	&22		&$-$	&\\  
  J000344$-$232355	&0.27051	&2	&42		&$-$	&\\
  J142249$-$272756	&0.27563	&1	&29		&$-$	&\\
  J042707$-$130253	&0.28929	&2	&122	&$-$	&\\
  J102837$-$010027	&0.32427	&2	&143	&$-$	&\\
  J110325$-$264515	&0.35896	&1	&31		&$-$	&\\
  J121140+103002	&0.39293	&1	&30		&$-$	&\\
  J050112$-$015914	&0.40310	&3	&103	&$-$	&\\
  J224752$-$123719	&0.40968	&2	&101	&$-$	&\\
  J220743$-$534633	&0.43720	&1	&26		&$-$	&\\  
  J044117$-$431343	&0.44075	&3	&60		&$-$	&\\
  J144653+011356	&0.44402	&1	&42		&$-$	&\\
  \hline
  \multicolumn{5}{c}{Other systems}\\
  \hline
  J044117$-$431343	&0.10114	&7	&130	&7.9	&g\\
  J231359$-$370446	&0.33980	&3	&41		&$-$	&\\
  J094253$-$110426	&0.39098	&2	&31		&$-$	&\\
  \hline
 \end{tabular}

 \medskip
 \textbf{References.} (a) \citet{brao}; (b) \citet{blane}; (c) \citet{blebrun}; \citet{bvladilo}; (e) \citet{bmiller}; (f) \citet{bjenkins}; (g) \citet{bchen}
\end{table*}
 
 \begin{figure}
  \resizebox{\hsize}{!}{\includegraphics{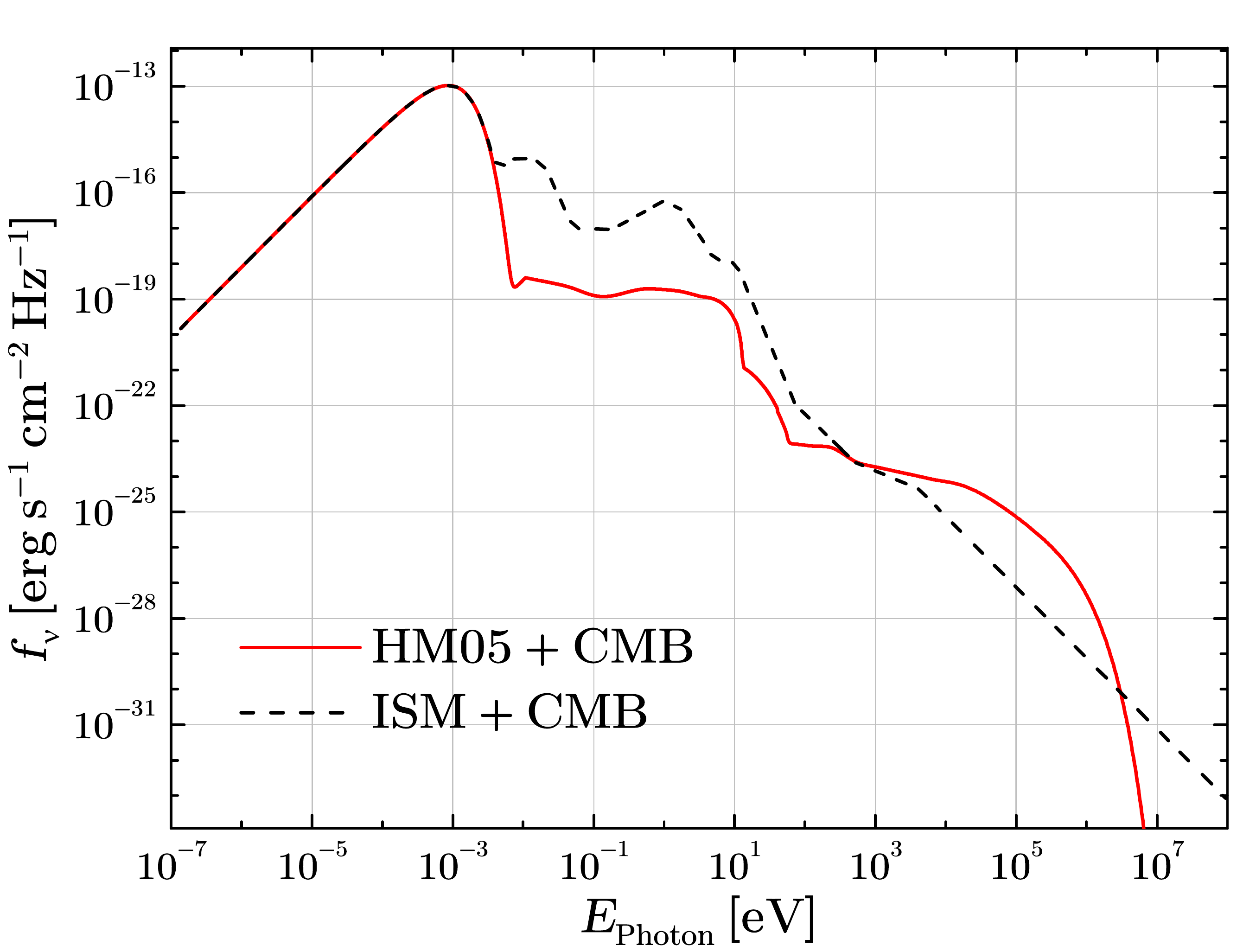}}
  \caption{Background radiation fields as described in the text.}
  \label{radfields}
\end{figure}
 
\begin{figure}
  \resizebox{\hsize}{!}{\includegraphics{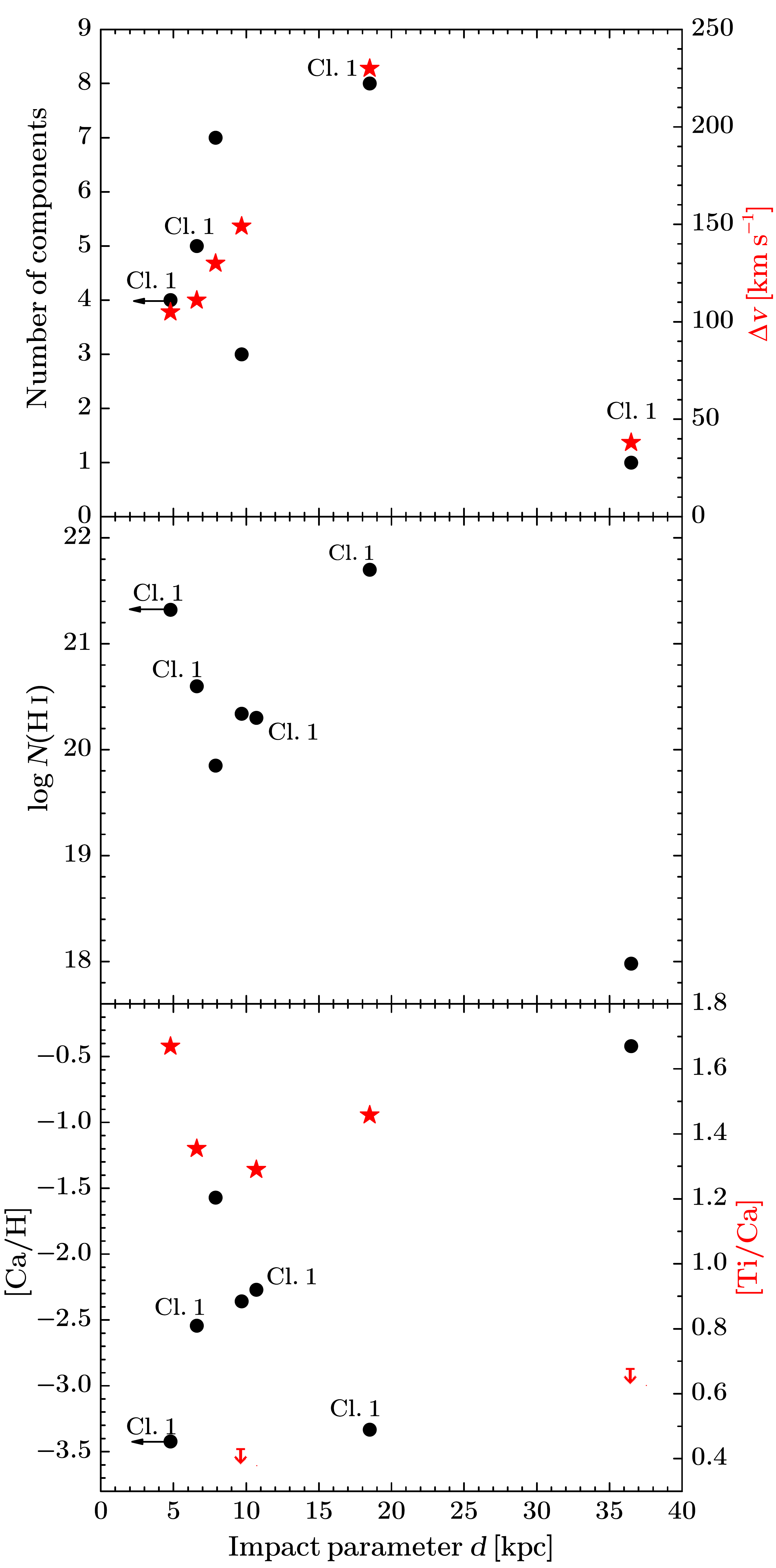}}
  \caption{Relations between $N_{\rm comp.}$, the number 
  of components, $\log N(\mathrm{H\,\textsc{i}})$, rest frame velocity width $\Delta v$ of 
  the Ca\,\textsc{ii} $\lambda3934$ absorption feature, ratios $[\mathrm{Ca}/\mathrm{H}]$, 
  $[\mathrm{Ti}/\mathrm{Ca}]$ and the impact parameter $d$. 
  Class 1 systems are labeled with ``Cl. 1''.}
  \label{allinall}
\end{figure}
 
\end{document}